\documentclass[11pt,onecolumn,draftcls]{IEEEtran}
\usepackage{latexsym}       
\usepackage{graphicx}       
\usepackage{rotating}       
\usepackage{subfigure}
\usepackage{epsfig}
\usepackage{times}
\usepackage{amssymb}
\usepackage{amsbsy}
\usepackage{amsmath}
\usepackage{epsf}

\newcommand{\bsh}{{\ensuremath{\hat{\mathbf s}}}}
\newcommand{\bst}{{\ensuremath{\tilde{\mathbf s}}}}
\newcommand{\lh}{{\ensuremath{\hat{l}}}}
\newcommand{\mh}{{\ensuremath{\hat{m}}}}

\newcommand{\ga}{\alpha}
\newcommand{\gah}{\mbox{$ \hat\alpha$}}
\newcommand{\gb}{\beta}
\newcommand{\grg}{\gamma}

\newcommand{\gth}{\theta}

\newcommand{\gm}{\mu}

\newcommand{\gr}{\rho}
\newcommand{\grh}{\mbox{$ \hat\rho$}}
\newcommand{\gs}{\sigma}

\newcommand{\gt}{\tau}

\newcommand{\gD}{\Delta}

\def\bm#1{\mbox{\boldmath $#1$}}
\newcommand{\vga}{\mbox{$\bm \alpha$}}
\newcommand{\vgah}{\mbox{{$\bm {\hat \alpha}$}}}

\newcommand{\mgG}{\mbox{$\bm \Gamma$}}

\newcommand{\bsigma}{{\bm\sigma}}

\newcommand{\rH}{^{ \raisebox{1pt}{$\rm \scriptscriptstyle H$}}}

\newcommand{\rE}{{\rm E}}

\newtheorem{theorem}{Theorem}[section]
\newtheorem{lemma}{Lemma}[section]

\newtheorem{prop}{Proposition}[section]
\newtheorem{claim}{Claim}[section]

\newtheorem{definition}{Definition}[section]

\newtheorem{question}{Question}[section]
\newtheorem{coro}{Corollary}[section]

\newcommand{\beq}{\begin{equation}}
\newcommand{\eeq}{\end{equation}}

\newcommand{\bea}{\begin{array}}
\newcommand{\ena}{\end{array}}
\newcommand{\bds}{\begin {itemize}}
\newcommand{\eds}{\end {itemize}}
\newcommand{\bdf}{\begin{definition}}
\newcommand{\blm}{\begin{lemma}}
\newcommand{\edf}{\end{definition}}
\newcommand{\elm}{\end{lemma}}
\newcommand{\bthm}{\begin{theorem}}
\newcommand{\ethm}{\end{theorem}}
\newcommand{\bprp}{\begin{prop}}
\newcommand{\eprp}{\end{prop}}
\newcommand{\bcl}{\begin{claim}}
\newcommand{\ecl}{\end{claim}}
\newcommand{\bcr}{\begin{coro}}
\newcommand{\ecr}{\end{coro}}
\newcommand{\bquest}{\begin{question}}
\newcommand{\equest}{\end{question}}

\newcommand{\rarrow}{{\rightarrow}}

\newcommand{\larrow}{{\larrow}}

\newcommand{\diag}{\ensuremath{\mathrm{diag}}}

\newcommand{\argmin}{\ensuremath{\mathrm{arg}\min}}
\newcommand{\argmax}{\ensuremath{\mathrm{arg}\max}}

\newcommand{\va}{{\ensuremath{\mathbf{a}}}}
\newcommand{\vab}{{\ensuremath{\mathbf{\bar{a}}}}}
\newcommand{\vah}{{\ensuremath{\mathbf{\hat{a}}}}}

\newcommand{\vb}{{\ensuremath{\mathbf{b}}}}

\newcommand{\ve}{{\ensuremath{\mathbf{e}}}}

\newcommand{\vh}{{\ensuremath{\mathbf{h}}}}

\newcommand{\vr}{{\ensuremath{\mathbf{r}}}}
\newcommand{\vs}{{\ensuremath{\mathbf{s}}}}
\newcommand{\vsh}{{\ensuremath{\mathbf{\hat{s}}}}}
\newcommand{\vst}{{\ensuremath{\mathbf{\tilde{s}}}}}

\newcommand{\vw}{{\ensuremath{\mathbf{w}}}}
\newcommand{\vwh}{{\ensuremath{\mathbf{\hat{w}}}}}

\newcommand{\vx}{{\ensuremath{\mathbf{x}}}}

\newcommand{\vzero}{\ensuremath{\mathbf{0}}}

\newcommand{\ba}{{\ensuremath{\mathbf{a}}}}

\newcommand{\br}{{\ensuremath{\mathbf{r}}}}
\newcommand{\bs}{{\ensuremath{\mathbf{s}}}}

\newcommand{\bw}{{\ensuremath{\mathbf{w}}}}
\newcommand{\bx}{{\ensuremath{\mathbf{x}}}}

\newcommand{\mA}{{\ensuremath{\mathbf{A}}}}

\newcommand{\mAko}{{\ensuremath{\mathbf{{A_k^{\odot}}}}}}

\newcommand{\mB}{{\ensuremath{\mathbf{B}}}}

\newcommand{\mC}{{\ensuremath{\mathbf{C}}}}

\newcommand{\mE}{{\ensuremath{\mathbf{E}}}}

\newcommand{\mH}{{\ensuremath{\mathbf{H}}}}

\newcommand{\mI}{{\ensuremath{\mathbf{I}}}}

\newcommand{\mJ}{{\ensuremath{\mathbf{J}}}}

\newcommand{\mM}{{\ensuremath{\mathbf{M}}}}

\newcommand{\mR}{{\ensuremath{\mathbf{R}}}}
\newcommand{\mRb}{{\ensuremath{\mathbf{\bar{R}}}}}
\newcommand{\mRh}{{\ensuremath{\mathbf{\hat{R}}}}}

\newcommand{\mS}{{\ensuremath{\mathbf{S}}}}

\newcommand{\mU}{{\ensuremath{\mathbf{U}}}}

\newcommand{\mV}{{\ensuremath{\mathbf{V}}}}

\def\IC{\mathbb C}
\def\IN{\mathbb N}
\def\IZ{\mathbb Z}
\def\IR{\mathbb R}

\def\shat{^{\mathchoice{}{}%
 {\,\,\smash{\hbox{\lower4pt\hbox{$\widehat{\null}$}}}}%
 {\,\smash{\hbox{\lower3pt\hbox{$\hat{\null}$}}}}}}

\def\bSigma{{
      \ooalign{
      \smash{\hskip.4pt\raise.4pt\hbox{$\Sigma$}}\vphantom{}\crcr
      \smash{\hskip.7pt\raise.6pt\hbox{$\Sigma$}}\vphantom{}\crcr
      \smash{\hbox{$\Sigma$}}\vphantom{$\Sigma$}}
      \vphantom{\hbox{$\Sigma$}}
      }}
\def\bTheta{{
      \ooalign{
      \smash{\hskip.5pt\raise.5pt\hbox{$\Theta$}}\vphantom{}\crcr
      \smash{\hskip.0pt\raise.1pt\hbox{$\Theta$}}\vphantom{}\crcr
      \smash{\hbox{$\Theta$}}\vphantom{$\Theta$}}
      \vphantom{\hbox{$\Theta$}}
      }}
\def\bDelta{{
      \ooalign{
      \smash{\hskip.4pt\raise.4pt\hbox{$\Delta$}}\vphantom{}\crcr
      \smash{\hskip.7pt\raise.6pt\hbox{$\Delta$}}\vphantom{}\crcr
      \smash{\hbox{$\Delta$}}\vphantom{$\Delta$}}
      \vphantom{\hbox{$\Delta$}}
      }}

\makeatletter

\def\bordermatrix#1{\begingroup \m@th
  \@tempdima 8.75\p@
  \setbox\z@\vbox{%
    \def\cr{\crcr\noalign{\kern2\p@\global\let\cr\endline}}%
    \ialign{$##$\hfil\kern2\p@\kern\@tempdima&\thinspace\hfil$##$\hfil
      &&\quad\hfil$##$\hfil\crcr
      \omit\strut\hfil\crcr\noalign{\kern-\baselineskip}%
      #1\crcr\omit\strut\cr}}%
  \setbox\tw@\vbox{\unvcopy\z@\global\setbox\@ne\lastbox}%
  \setbox\tw@\hbox{\unhbox\@ne\unskip\global\setbox\@ne\lastbox}%
  \setbox\tw@\hbox{$\kern\wd\@ne\kern-\@tempdima\left[\kern-\wd\@ne
    \global\setbox\@ne\vbox{\box\@ne\kern2\p@}%
    \vcenter{\kern-\ht\@ne\unvbox\z@\kern-\baselineskip}\,\right]$}%
  \null\;\vbox{\kern\ht\@ne\box\tw@}\endgroup}
\makeatother

\newcommand{\DL}{\begin{dashlist}}
\newcommand{\DLE}{\end{dashlist}}

\makeatletter
\def\argmin{\mathop{\operator@font arg\,min}}
\def\argmax{\mathop{\operator@font arg\,max}}
\makeatother

\newcommand{\SI}{\begin{indlist}}
\newcommand{\EI}{\end{indlist}}

\newcommand{\lt}{{\ensuremath{\tilde{l}}}}
\newcommand{\mt}{{\ensuremath{\tilde{m}}}}

\newcommand{\mRi}{\mR^{-1}}
\newcommand{\dfdl}{\frac{\partial f(\bs_i, \mR^{-1})}{\partial l}}
\newcommand{\dfdm}{\frac{\partial f(\bs_i, \mR^{-1})}{\partial m}}
\newcommand{\dfTdlT}{\frac{\partial^2 f(\bs_i, \mR^{-1})}{\partial l^2}}
\newcommand{\dfTdmT}{\frac{\partial^2 f(\bs_i, \mR^{-1})}{\partial m^2}}
\newcommand{\dfTdldm}{\frac{\partial^2 f(\bs_i, \mR^{-1})}{\partial l\partial m}}
\newcommand{\dfTdmdl}{\frac{\partial^2 f(\bs_i, \mR^{-1})}{\partial m\partial l}}
\newcommand {\mul}{\ba^H\mR^{-1}\ba}

\newcommand {\dadlRdadl}{\frac{\partial \ba^H}{\partial l}\mR^{-1}\frac{\partial \ba}{\partial
l}}

\newcommand {\aRdaTdlT}{\ba^H\mR^{-1}\frac{\partial ^2 \ba}{\partial l^2}}

\newcommand {\aRdadl}{\ba^H\mRi\frac{\partial \ba}{\partial l}}

\newcommand {\aRdadm}{\ba^H\mRi\frac{\partial \ba}{\partial m}}

\begin{document}
\title{Parametric high resolution techniques for radio astronomical imaging}
\author{Chen Ben-David and Amir Leshem {\em (Senior Member)} \\
    School of Engineering\\
    Bar-Ilan University\\
    52900, Ramat-Gan\\
    Email: leshema@eng.biu.ac.il
}
\date{Submitted: Feb. 1st 2008, Revised: June 11th 2008.}
\maketitle
\begin{abstract}
    The increased sensitivity of future radio telescopes will result in requirements for
    higher dynamic range within the image as well as better resolution and immunity to interference.
    In this paper we propose a new matrix formulation of the imaging equation in the cases of non co-planar arrays and polarimetric
    measurements.
    Then we improve our parametric imaging techniques in terms of resolution and estimation accuracy. This is done by enhancing both the
    MVDR parametric imaging, introducing alternative dirty images and by
    introducing better power estimates based on least squares, with positive semi-definite constraints. We also discuss the use of robust
    Capon beamforming and semi-definite programming for solving the self-calibration problem. Additionally we provide statistical analysis of the bias of the MVDR beamformer for the case of moving array, which serves as a first step in analyzing iterative approaches such as CLEAN and the techniques proposed in this paper.
     Finally we demonstrate a full deconvolution process based on the parametric imaging
     techniques and show its improved resolution and sensitivity compared to the CLEAN method.
\end{abstract}
\section{Introduction}
The future of radio astronomical discoveries depends on achieving
better resolution and sensitivity while maintaining immunity to
terrestrial interference which is rapidly growing. To achieve the
improved sensitivity and higher resolution new instruments are being
designed. The Square Kilometer Array (SKA) \footnote{For information on the SKA project the reader is
referred to
http:$//$www.skatelescope.org.} \cite{hall2005} and the Low Frequency
Array (LOFAR) \cite{bregman2005}
are two of these advanced instruments. Achieving
higher sensitivity to observe faint objects results in  high dynamic
range requirements within the image, where strong sources can affect
the imaging of the very weak sources. On the other hand, Moore's law
\cite{moore65} together with recent advances in optimization theory
\cite{boyd2004} open the way to the application of more advanced
computational techniques. In contrast to hardware implementation, these image formation algorithms,
that are implemented in software can benefit from the continuing increase in computational power,
even after the antennas and the correlators will be built. In this paper we extend the parametric
deconvolution approach of \cite{leshem2000a} to obtain better power
estimation accuracy and higher robustness to interference and
modeling uncertainty. The algorithms presented here can also be used
in conjunction with real-time interference mitigation techniques as
described in \cite{leshem2000a} and \cite{leshem2000b}.

We briefly describe the current status of radio astronomical imaging
techniques. For a more extensive overview the reader is referred to
\cite{thompson86}, \cite{yen85} or \cite{taylor99}. A good historic
perspective can be found in \cite{kellerman2001}, whereas
\cite{sault2007} provides a very recent perspective.

The principle of radio interferometry has been used in radio astronomy since 1946 when Ryle and Vonberg
constructed a radio interferometer using  dipole antenna arrays \cite{ryle52}. During
the 1950's several radio interferometers which use the synthetic aperture
created by movable antennas have been constructed. In 1962 the
principle of aperture synthesis using earth rotation has been proposed
\cite{ryle62}. The basic idea is to exploit the rotation of the earth to
obtain denser coverage of the visibility domain (spatial Fourier domain).
The first instrument to use this principle was the five kilometer
Cambridge radio telescope.  During the 1970's new instruments with
large aperture have been constructed. Among these we find the Westerbork
Synthesis Radio Telescope (WSRT) in the Netherlands and the Very Large
Array (VLA) in the USA. Recently, the Giant Microwave Telescopes (GMRT) has been constructed in India
and the Allen Telescope Array (ATA) in the US. Even these instruments subsample the Fourier
domain, so that unique reconstruction is not possible without some
further processing known as deconvolution. The deconvolution process
uses some a-priori knowledge about the image to remove the effect of
``dirty beam'' side-lobes.

Two principles dominate the astronomical imaging deconvolution. The
first method was proposed by Hogbom \cite{hogbom74} and is known as
CLEAN. The CLEAN method is basically a sequential Least-Squares (LS)
fitting procedure in which the brightest source location and power
are estimated. The response of this source is removed from the image
and then the process continues to find the next brightest source,
until the residual image is noise-like.  During the years it has
been partially analyzed \cite{schwarz78}, \cite{schwarz79} and
\cite{tan86}. However full analysis  of the method is still lacking
due to its iterative nature. The CLEAN algorithm has many recent
flavors which are capable of faster performance e.g., the Clark
version \cite{clark80} and the Cotton-Schwab \cite{schwab84}. In
these versions the dirty image is recomputed only after several
point sources have been estimated. Furthermore the sources are
subtracted from the ungridded visibility. This results in better
suppression of the sources. Interestingly it is well known that when the noise model is non-white
(as in radar clutter modeled by ARMA processes in SAR applications) algorithms such as RELAX \cite{li96}
outperform the CLEAN algorithm.

A second approach proposed by Jaynes \cite{Jaynes57} is
maximum entropy deconvolution (MEM). The basic idea behind MEM is
the following.  Not all images which are consistent with the measured
data and the noise distribution satisfy the positivity demand,
i.e., the sky brightness is a positive function. Consider only those
that satisfy the positivity demand.
  From these select the one that is most likely to have been created randomly.
This idea has also been proposed in \cite{frieden72} and applied to
radio astronomical imaging in \cite{gull78}. Other approaches based
on the differential entropy have also been proposed \cite{ables74}, \cite{wernecke77}. An extensive collection of papers discussing the various
methods and aspects of maximum entropy can be found in the various papers in
\cite{roberts84}.
Briggs \cite{briggs95} proposed a non-negative least squares approach (NNLS) which eliminates the
need for iterative processing. However, the computational complexity is very large.

In this paper we use a reformulation of the image formation problem
as a parameter estimation problem using a set of covariance
matrices, measured at the various observation epochs
\cite{leshem2000a}. This yields a model where the array response is
time varying. Previous research on time varying arrays and their
application to direction-of arrival (DOA) estimation includes
\cite{zeira95}, \cite{zeira96} and \cite{sheinvald98}. In \cite{leshem2000a} we proposed a simplified
ML estimator. Lanterman \cite{lanterman2000} developed a full EM
algorithm for implementing the MLE proposed in \cite{leshem2000a}.
The algorithm performs quite well although it is quite complex
compared to the solutions described in this paper.

The above mentioned algorithms assume perfect knowledge of the
instrumental response (point spread function). Due to various internal
and external effects this assumption holds only approximately. One way
to overcome this problem is the use of calibrating sources. An unresolved
source with known parameters is measured, and by relating the model errors
to the array elements a set of calibration equations is solved. A much
more appealing solution is to try to improve the fitting between the
data and the sky model by adjusting the calibration parameters. Another
possibility \cite{noordam82} is to use the redundant structure of the
array to solve for the calibration parameters (this is possible only
for some arrays which have redundant baselines, such as the WSRT). A good
overview of the various techniques is given in \cite{pearson84}.

In this paper we extend the above methods in several directions.
First, we extend the parametric formulation of \cite{leshem2000a} to the non-coplanar array case and to
polarimetric imaging.
Then we propose relatively ``low'' complexity approaches for the
image deconvolution problem, based on Minimum Variance
Distortionless Response (MVDR) and its robust extensions. We call these extensions Least-Squares Minimum Variance Imaging (LS-MVI). We provide
a new type of dirty image that has isotropic noise response,
something desirable in imaging applications. This is done by
generalizing the work of Borgiotti and Kaplan \cite{borgiotti79} to
the moving array case. We discuss acceleration techniques, related
to the Clark \cite{clark80} and the Cotton-Schwab \cite{schwab84}
approaches to CLEAN. However, in contrast to the classical CLEAN
case, the accelerated algorithm involves semi-definite programming
of low order, ensuring that the covariance matrices remain positive
semi-definite after the subtraction. We provide analytic expressions
for the asymptotic bias of the MVDR based imaging. We also relate
the classical self-calibration technique to a novel extension of the
robust Capon beamformer to the moving array case, showing that
self-calibration can be cast in terms of semi-definite programming.
This has advantage over previous approach to self-calibration since
the covariance structure maintains its positive definite structure.

We also demonstrate full parametric deconvolution process based on
the proposed technique and compare CLEAN and LS-MVI on simulated
images. We will show in simulations that LS-MVI, indeed,
significantly outperforms the classical CLEAN algorithm over a wide
range of parameters, in terms of better resolution, higher
sensitivity and dynamic range.

The structure of the paper is as follows.  In section
\ref{sec:astron} we describe the astronomical measurement equation.
The measurement equation is subsequently rephrased in a more
convenient matrix formulation both for non co-planar and co-planar arrays. It is then extended with the effect of noise and unknown calibration parameters. We also discuss extension to polarimetric measurements. In
section \ref{sec:LS_MVI} we discuss the new Least Squares Minimum
Variance Imaging and extensions to self calibration using robust Capon techniques. In section \ref{sec:MVDR_analysis} we present bias
analysis of the MVDR imaging technique.

In \ref{sec:simulation} we describe several computer simulations demonstrating the
 gain in parametric deconvolution relative to the CLEAN in terms of
resolution, sensitivity and capability to model extended structures. We also provide example of the tightness of the statistical bias analysis of the MVDR DOA estimation with a moving array.
 We end up with some conclusions.

\section{Astronomical measurement equations}
\label{sec:astron}
    In this section we describe a simplified mathematical model for the
    astronomical measurement and imaging process.  Our discussion follows
    the introduction in \cite{taylor99} and its matrix extension in \cite{leshem2000a}.
    We extend the matrix formulation of \cite{leshem2000a} to non co-planar arrays and polarimetric measurements.
    This allows us to obtain a uniform description of various astronomical imaging operations such as
    deconvolution and self-calibration.
\subsection{The interferometric measurement equation}
    The waves received from the celestial sphere may be considered as
    spatially incoherent wideband random noise. They may be polarized and can
    contain spectral absorption or emission lines.
    Rather than considering the emitted electric field at a location on
    the celestial sphere, astronomers try to recover the {\em
    intensity} (or brightness) $I_f(\vs)$ in the direction of
    unit-length vectors $\vs$, where $f$ is a specific frequency.
    Let $E_f(\vr)$ be the received celestial electric field at a location
    $\vr$ on earth and $A(\vs)$ be the antenna gain towards direction $\vs$.
    The measured correlation of the electric fields
    between two identical sensors $i$ and $j$ with locations $\vr_i$ and $\vr_j$
    is called {\em visibility} and is (approximately) given by
    \cite{taylor99}\footnote{To simplify notation, from this point we do not include the directional
    response of the elements of the radio telescope.}
\beq
    V_f(\vr_i,\vr_j)
    :=    \rE\{ E_f(\vr_i) \overline{E_f(\vr_j)} \}
    \;=\; \int_{\mbox{sky}}
          A(\vs)I_f(\vs) e^{-2\pi \jmath f\; \bs^T(\vr_i - \vr_j)/c}\; d\Omega
    \,.
\eeq
    $\rE\{\,\cdot\,\}$ is the mathematical expectation operator,
    the superscript $^T$ denotes the transpose of a vector, and overbar denotes
    the complex conjugate.
    Note that it is only dependent on the oriented distance $\vr_i-\vr_j$
    between the two antennas; this vector is called a {\em baseline}.

    For simplification, we may sometimes
    assume that the astronomical sky is a collection of $d$
    discrete point sources (maybe unresolved). This gives
\[
    I_f(\vs) = \sum_{l=1}^d I_f(\vs_l) \delta(\vs - \vs_l) \,,
\]
    where $\vs_l$ is the coordinate of the $l$'th source, and thus
\begin{equation}
\label{eq:discretesource}
    V_f(\vr_i,\vr_j) \;=\;
      \sum_{l=1}^d
        \,
        I_f(\vs_l)
        \,
        e^{-2\pi \jmath f\, \vs_l^T(\vr_i - \vr_j)/c}  \,.
\end{equation}
    For earth rotation synthesis arrays, the following coordinate system is often
    used.  We assume an array with antennas that
    have a small field of view and that track a reference source
    direction $\bs_0$ in the sky.
    Other locations in the field of view can be written as
$
    \bs = \bs_0 + \bsigma\,, \bs_0 \perp \bsigma
$
    (valid for small $\bsigma$) and a natural coordinate system is
$
       \bs_0 = [0,\, 0,\, 1]^T, \bsigma = [\ell,\, m,\, 0]^T \,.
$
    Similarly, the receiver baselines can be
    parameterized as
$
    \br_i - \br_j = \lambda [u,\, v,\, w]^T \,,
    \qquad \lambda = \displaystyle\frac{c}{f} \,.
$
    The measurement equation in $(u,v,w)$ coordinates thus becomes (see Perley \cite{taylor99} chapter 19)
\beq \label{noncoplanar}
V(u,v,w)=\int_{-\infty}^{\infty}\int_{-\infty}^{\infty}\frac{I(l,m)}{\sqrt{1-l^2-m^2}}e^{-2\pi
j \left(ul+vm+w (\sqrt{1-l^2-m^2}-1)\right)}dl dm \eeq Assuming
phase tracking is performed to compensate for the geometric delay we
obtain \beq \label{noncoplanar1}
V'(u,v,w)=\int_{-\infty}^{\infty}\int_{-\infty}^{\infty}\frac{I(l,m)}{\sqrt{1-l^2-m^2}}
e^{-2\pi j \left(ul+vm+wn \right)}dl dm \eeq where $n=
\sqrt{1-l^2-m^2}$. Assuming that the sky is composed of discrete set
of sources we obtain \beq \label{eq:discretesource_ncp}
V'(u,v,w)=\sum_{\ell=1}^d
\frac{I(l_\ell,m_\ell)}{\sqrt{1-l_\ell^2-m_\ell^2}} e^{-2\pi j
\left(ul_\ell+vm_\ell+w n_\ell \right)} \eeq In certain conditions,
such as East-West array or when the field of view is limited by the
antenna primary beam to sufficiently small angular region, the third
term in the exponential can be neglected and the term
$\sqrt{1-l_\ell^2-m_\ell^2}$ is approximately $1$. The measurement
equation (\ref{noncoplanar}) becomes:
\begin{equation}                    \label{eq:fourier}
    V_f(u,v)
    \;=\;
      \int\!\!\!\int\;
      I_f(l,m)\, e^{-2 \pi \jmath(ul + v m)}\, dl d m
    \,.
\end{equation}
    which has the form of a Fourier transformation.

    The function $V_f(u,v)$ is sampled at various coordinates $(u,v)$
    by first of all taking all possible sensor pairs $i,j$ or baselines
    $\br_i-\br_j$, and second by realizing that the sensor locations
    $\br_i$, $\br_j$ are actually time-varying since the earth
    rotates.  Given a sufficient number of samples in the $(u,v)$ domain, the
    relation can be inverted to obtain an image (the `map').
    Direct Fourier inversion of the visibility data suffers from severe aliasing and is termed the dirty image. To       overcome the aliasing, deconvolution algorithms are required.
\subsection{Matrix formulation of the measurement equation}       \label{sec:array}
    To allow a parametric imaging formulation of the discrete source model, we
    can now formulate our measurement equations in terms of matrices following \cite{leshem2000a}. Our formulation
    is more general then \cite{leshem2000a}, extending it to the non co-planar array case.

    Let $\vr_0(t_k)$ be an arbitrary and time-varying
    reference point, typically at one of the elements
    of the array, and let us take the $(u,v,w)$ coordinates of the other
    telescopes with respect to this reference,
$
   \br_i(t) - \br_0(t) \;=\; \lambda [u_{i0}(t),\; v_{i0}(t),\; w_{i0}(t)] \,,
   \qquad
   i = 1, \cdots, p\,.
$
    Similarly to \cite{leshem2000a} in the coplanar case, equation (\ref{eq:discretesource_ncp}) can then
    be written in matrix form as
\begin{equation}
\label{v_mtx}
   \mR_k \;=\; \mA_k  \mB \mA\rH_k,
\end{equation}
    where
\[
    \mR_k \equiv \mR(t_k),\quad \mA_k = \left[\va_k(l_1,m_1),\ldots,\va_k(l_d,m_d)\right]\,,
\]
\begin{equation}
\label{eq:vadef}
    \va_k(l,m) =
    \left[ \bea{c}
    e^{-2\pi \jmath(u_{10}(t_k) l + v_{10}(t_k) m +w_{10}(t_k) n)} \\
    \vdots \\
    e^{-2\pi \jmath(u_{p0}(t_k) l + v_{p0}(t_k) m +w_{p0}(t_k) n)}
    \ena \right]
\end{equation}
and
\[
\mB=\diag\left[\frac{I(l_1,m_1)}{\sqrt{1-l_1^2-m_1^2}},\ldots, \frac{I(l_d,m_d)}{\sqrt{1-l_d^2-m_d^2}} \right]\,.
\]

    When the field of view is limited the factor $\sqrt{1-l_\ell^2-m_\ell^2}$ is close to unity and can be neglected. The vector function $\va_k(l,m)$ is called
    the {\em array response vector} in array signal processing.  It
    describes the response of the telescope array
    to a source in the direction
    $(l,m)$. As usual, the array response is frequency dependent.
    The response is also slowly time-varying due to
    earth rotation.  It is assumed that the function as shown
    here is completely known. However, typically the array response is not perfectly known.
    Each antenna may have a different complex
    receiver gain, $\gamma_i(t)$, dependent on cable losses, amplifier gains, and (slowly) time varying.
    For LOFAR type arrays the calibration is also space varying because of
    atmospheric conditions. To simplify the exposition we only treat calibration parameters that are
    spatially invariant assuming that the field of view is contained in a single isoplanatic patch. The case of LOFAR type array with space varying calibration parameters is complicated, and will be treated in a subsequent paper. The measurements are also containing additive system noise.
    Typically this noise is zero mean, independent among
    the antennas (thus spatially white). After noise calibration and scaling of the measurements
    we can also assume that
    it has a covariance matrix that is a multiple of the identity matrix,
    $\sigma^2 \mI$, where $\sigma^2$ is the noise power on a single antenna.
    Usually the receiver noise is assumed to be Gaussian.
    The resulting model of the received covariance matrix then becomes
\begin{equation}
\label{self_cal_model}
    \mR_k \;=\; \mgG_k \mA_k \mB \mA\rH_k \mgG_k\rH  \;+\;  \gs^2 \mI
\end{equation}
    where
\begin{equation}
\label{def_Gammak}
    \mgG_k =\diag
    \left[\gamma_{1,k},\ldots,\gamma_{p,k} \right] \,
\end{equation}
and $\gamma_{i,k}=\gamma_i(t_k)$ is the calibration parameter for antenna $i$ at time epoch $t_k$.
To allow proper calibration it is assumed that the time varying calibration parameters vary slowly so calibration equations can combine multiple epochs.
\subsection{The astronomical measurement equation for polarized sources}
Next, we extend our matrix formulation to the case of polarimetric measurement of polarized sources.
Our notation follows the conventions of Hamaker et. al \cite{hamaker96}-\cite{hamaker2000}.
For concreteness, we concentrate on circular feeds. We perform the deconvolution of
the cross-polarization
parameters and then recover the Stokes parameters using the Muller matrices for each source.
This enables a straightforward
extension of the matrix formulation to polarized sources. First we restrict our attention to purely
orthogonal feeds, and then
introduce the Jones matrices of each antenna into the model.
To that end assume that each antenna has two orthogonal circular feeds.
As usual we restrict our discussion to the quasi-monochromatic case.
The response of the array at epoch $k=1,...,K$
towards direction $\vs$ can now be decomposed as
\beq
\left[\va_{L,k}(\vs), \va_{R,k}(\vs)\right]=\va_k(\vs)\otimes \mI_2
\eeq
where $\mI_2$ is the 2x2 identity matrix. This implies that for any directions $\vs_1,\vs_2$ the vectors
$\va_{L,k}(\vs_1),\va_{R,k}(\vs_2)$ are orthogonal.
Assume that the sky is composed of $d$ point sources (to replace the integral by a finite sum).
The extended measurement equation for ideal feeds is now described by
\beq
\label{polcal_equation}
\mR_k=\mAko \mB \left(\mAko\right)^H + \gs^2 \mI
\eeq
where
\beq
\mAko =
\left[
\va_{L,k}(\vs_1),\va_{R,k}(\vs_1),...,\va_{L,k}(\vs_d) \va_{R,k}(\vs_d)
\right]
\eeq
is the array response matrix,
\beq
\mB=\diag \left\{\mE_1,\ldots,\mE_d \right\}
\eeq
is a block diagonal matrix with 2x2 sub-blocks, $\mE_\ell$, on the diagonal consisting of the source
coherency matrices
\beq
\mE_\ell = \left[
\bea{cc}
\langle e_{L,\ell} e_{L,\ell}^* \rangle & \langle e_{L,\ell} e_{R,\ell}^* \rangle \\
\langle e_{R,\ell} e_{L,\ell}^* \rangle & \langle e_{R,\ell} e_{R,\ell}^* \rangle \\
\ena \right]. \eeq Similarly to the calibrated model for scalar
imaging, we can introduce feed calibration matrices $\mgG_k$ for
non-ideal (non-orthogonal) feeds, using the Jones matrices:
\beq
\mgG= \diag \left\{\mJ^1,\ldots,\mJ^p\right\},
\eeq
where the 2x2 sub-blocks are the corresponding Jones matrices representing the
polarization leakage and the different gains and phases of the
feeds:
\beq \mJ^\ell=\left[
\bea{cc}
J^\ell_{11} & J^\ell_{12} \\
J^\ell_{21} & J^\ell_{22}
\ena \right].
\eeq
The calibrated measurement equation in matrix form now becomes
\beq \mR_k=\mgG_k \mAko \mB \left(\mAko\right)^H\mgG_k^H + \gs^2 \mI.
\eeq
This model is very similar to the scalar measurement equation
(\ref{self_cal_model}), except that the source matrix is now block
diagonal. This will have an effect on the parametric imaging
algorithms. Note that it is easy to see that the 2x2 sub-blocks of
the matrix $\mR_k$ agree with the formulation of Hamaker et. al
\cite{hamaker2000}. After deconvolution of the model
(\ref{polcal_equation}) we can extract the Stokes parameters of each
source using the Muller matrices $\mS$
\beq \vb_\ell=\mS \ve^S_\ell,
\eeq
where $\vb_\ell=\hbox{vec}\left({\mE_\ell}\right)$ and
\beq
\ve^S= \left[I_\ell, Q_\ell, U_\ell, V_\ell \right]^T.
\eeq
\section{The Least-Squares Minimum-Variance Imaging}
\label{sec:LS_MVI} The idea of using Direction-of-Arrival (DOA)
estimation techniques for imaging was first introduced in
\cite{leshem2000a}. In that paper it was suggested that the imaging
process will be based on MVDR (Minimum Variance Distortionless
Response) DOA estimates \cite{capon69}, using classical dirty image
intensities, similar to the CLEAN method \cite{hogbom74}. Therefore,
the algorithm improves the source location estimates, but the power
estimates are still inaccurate similarly to the CLEAN algorithm.
In this paper we improve the algorithm of \cite{leshem2000a} in
several directions: First, we introduce a new type of dirty image
that has better properties than the standard MVDR dirty image. We
term this new dirty image the Adaptive Angular Response (AAR) dirty
image, since it extends the technique of \cite{borgiotti79} to the
moving array case. The main
advantage is the isotropic noise response, similar to that of the
classical Fourier beamformer. Then, we improve the power estimate
using non-negative least squares estimator for the power and
imposing a positive semi-definite constraint on the residual covariance
matrices. Note that in contrast to the global NNLS of Briggs
\cite{briggs95} we only solve linear NNLS which has a closed form
solution. Furthermore, the dimensionality of the specific NNLS used
in this paper is very low, operating on a single (or few) sources at
a time. The two previous papers on parametric imaging
\cite{leshem2000a}, \cite{leshem2004} have only demonstrated dirty
images based on MVDR and the Robust MVDR estimation. In this paper
we demonstrate the benefits of the parametric
approach after a full deconvolution. The proposed algorithm results
in a robust technique capable of much better resolution compared to
the CLEAN, since it inherently suppresses interference from other
sources within the image, by using a data dependent beam in order to
estimate the locations. Finally, we will show how to speed up the
proposed algorithm by estimating multiple sources at each round.

The MVDR based imaging can be generalized to Robust Capon
Beamforming (RCB)
\cite{leshem2004,stoica03,li03,vorobyov03,lorenz05, verdu84}, which
is also expected to provide robustness to array manifold errors. A
first stage of experimental verification of using the RCB has been
carried out in \cite{leshem2004}, where dirty images based on the
RCB showed superior interference immunity compared to the ordinary
dirty images. We also demonstrate how to combine the LS-MVI
with the self-calibration. However, this requires a novel extension
of the robust Capon algorithm to the case of moving array using
semi-definite programming. We begin with discussion of the MVDR and
the robust MVDR algorithms for a moving array and then discuss power
estimation. Then we provide a complete description of the algorithm
and its accelerated versions. We end up with extending the robust
Capon beamformer to the moving array case and demonstrate how it can
be used for self calibration.
\subsection{Two dimensional minimum variance  estimation with a moving array} \label{sec:mvdr}
The Minimum Variance Distortionless Response
method ( MVDR ) \cite{capon69} was one of the first super-resolution techniques
for direction-of-arrival estimation. Compared to classical (Fourier) based beamforming it provides better
separation of closely spaced sources and robustness to strong interference. The MVDR estimator is obtained by solving
the following optimization problem \cite{johnson93}:
\newline
Let $\vx(t_k)$ denote the antennas' output signal at time $t_k$ (we assume that DC has been removed so that
$\vx(t_k)$ is a zero mean random signal). We
apply a weight vector $\bw_k$ to $\vx(t_k)$.
Hence, the variance of the antennas' output, derived from pointing
towards direction $\bs$, is given by:
\[
Var[\bw_k^H(\bs)\bx(t_k)]= \bw_k^H(\bs)\mRh_k\bw_k(\bs)\,.
\]
The total output power is then given by
\begin{equation}
\label{mvdr_image}
    I'_D(\bs):=\sum_k \bw_k^H(\bs)\mRh_k\bw_k(\bs)\,.
\end{equation}
Note that for the classical dirty image $\bw_k(\bs)=\ba_k(\bs)$ \cite{leshem2000a} and the dirty
image is given by
\begin{equation}
\label{def_dirty_image}
    I_D(\bs):=\sum_k \ba_k^H(\bs)\mRh_k\ba_k(\bs)\,.
\end{equation}
This amounts to using fixed Fourier basis vectors for beamforming,
independently of the data. As in \cite{leshem2000a}, we can replace
$\mRh$ by $\mRh-\gs^2\mI$, where $\gs^2$ is the noise power. This
power downdating can, alternatively, be incorporated into the
positive semidefinite constraints. A much better approach to the
imaging process would be to minimize the interference subject to
transferring the desired direction unchanged. This is equivalent to
working with data dependent beamformer. Using the data to form the
beam provides much better interference suppression, compared to the
fixed Fourier basis. The variance of the array output consists of
the response to sources at many directions. We require that at each
time instance $k$ the total output power will be minimized, subject
to the constraint that the output response towards direction $\bs$
will be fixed. This is equivalent to requiring that the
contributions of the sidelobes of other sources will be minimized.
Using equation (\ref{self_cal_model}) and assuming calibrated array
we obtain
\[
    \mR_k=\mA_k \mB\mA_k^H+\gs^2 \mI\,.
\]
In order to obtain an estimate of the power originating from direction $\bs$ without
interference, we require that for all $k$
\begin{equation}
    \bw_k^H(\bs)\ba_k(\bs)=1
\end{equation}
and then minimize the overall output power. This can be reformulated as
the following constrained problem (For simplicity we denote
$\bw_k(\bs)$ by $\bw_k$):
\begin{equation}
    \vwh_k(\vs) \;=\; \argmin_{\vw_k}\; \vw_k\rH \mRh_k \vw_k
    \qquad\hbox{\ subject to \ } \quad
   \vw_k\rH\va_k(\vs) = 1 \,.
\end{equation}
The above problem can be solved using Lagrange multipliers.
The solution is given by:
\begin{equation}
    \vwh_k(\bs) = \gb_k(\bs) \mRh_k^{-1} \va_k(\vs)
    \,,
    \qquad\mbox{where } \quad
    \gb_k(\bs) = \frac{1}{\va_k\rH(\vs) \mRh_k^{-1} \va_k(\vs)} \,.
\end{equation}
Substituting $\vwh_k(\bs) $ in Eq. (\ref{mvdr_image}) we obtain
\begin{equation}
\label{mvdrImage1}
    \mI_D'(\bs) \;=\;
       \sum_{k=1}^K \;\frac{1}{\va_k\rH(\vs) \mRh_k^{-1} \va_k(\vs)}
    \,.
\end{equation}
\begin{definition}
\label{def:mvdr_dirty_image}
    $\mI_D'(\bs)$ formulated in Eq. (\ref{mvdrImage1}) is the {\textit{MVDR dirty
    image}}.
\end{definition}

Finally, we describe a second variant of the MVDR dirty image, which combines the adapted angular response
(AAR) of \cite{borgiotti79} and the approach of \cite{rieken2004} to moving arrays.
Borgiotti and Kaplan \cite{borgiotti79} proposed to use an MVDR type of estimator
$\vw(\vs)=\gm \mRh^{-1} \va(\vs)$ but in order to obtain isotropic behavior of the noise to
add a constraint $\|\vw(\vs) \|^2=1$ for all $\vs$. This results in the following spatial spectrum estimator
\beq
\label{AAR}
\frac{\va(\vs)^H \mRh^{-1}\va(\vs)}{\va(\vs)^H \mRh^{-2}\va(\vs)}.
\eeq
Rieken and Fuhrmann \cite{rieken2004} suggested to assume that the measurements at each epoch
(or array in their formulation) are
uncorrelated, resulting in block diagonal covariance matrix
\beq
\label{def_mRb}
\mRb=\diag \left\{\mRh_1,\ldots,\mRh_k\right\}.
\eeq
This assumption holds for the radio-astronomical case, since the received signal can be assumed independent
over different time epochs. Let
\beq
\label{def_vab}
{\bar \vw}=\gm \mRb^{-1} \vab,
\eeq
where,
\[
\vab=\left[\va_1(\vs)^T,...,\va_K(\vs)^T \right]^T.
\]
Substituting (\ref{def_mRb}) and (\ref{def_vab}) into (\ref{AAR}),
we obtain that the dirty image is given by \beq \label{AAR_dirty}
I_D^{''}(\vs)=\frac{\sum_{k=1}^K \va_k(\vs)^H
\mRh_k^{-1}\va_k(\vs)}{\sum_{k=1}^K\va_k(\vs)^H
\mRh_k^{-2}\va_k(\vs)}\,. \eeq This new spatial power spectrum
estimator for the moving array case has isotropic white noise
response, and optimal suppression of interference under the
isotropic white noise requirements. Furthermore, the averaging over
the time epochs $k=1,...,K$ at both numerator and denominator
results in smoother behavior of the dirty image.
\begin{definition} \label{def:AAR_dirty_image}
    $\mI_D''(\bs)$ formulated in Eq. (\ref{AAR_dirty}) is the {\textit{AAR dirty
    image}}.
\end{definition}
The AAR dirty image (\ref{AAR_dirty}) yields better power estimates and is more robust to strong
noise, but has somewhat higher computational complexity.
\subsection{The LS-MVI Algorithm} \label{sec:MVI}
In this section we present a new algorithm - The Least Squares
Minimum Variance Imaging (LS-MVI). The algorithm is an iterative
one, similar to the CLEAN method. 
We assume that every cosmic source in the sky brings the MVDR dirty
image (\ref{mvdrImage1}) or the AAR dirty image (\ref{AAR_dirty}) to its maximum at its direction.
Thus, in each iteration we find the brightest point in the MVDR
dirty image. Then we estimate its intensity, using Least Squares, as
described below. We subtract part of the source's contribution to
the correlation matrices. A new MVDR dirty image is then calculated,
using the new correlation matrices. We continue the iterations until
a certain stopping rule (typically defined by $\chi^2$ test for the
residual dirty image) is met. The final image is composed of the
locations and intensities we have found during these iterations,
convolved with a synthesized beam, usually an ideal Gaussian beam.
Note that the location estimator is not limited to the grid of the dirty image. Since the model is continuous
either interpolation or local optimization can be used to find the best location, independent of the grid. Typically,
quadratic interpolation around the maximum of the grid suffices. This solves the dynamic range problem pointed out
by \cite{voronkov2004}.
In \cite{leshem2000a} the locations of the sources were estimated by using the MVDR dirty image $I_D'(\vs)$.
However, the intensities were estimated using the conventional dirty image $I_D(\vs)$.
As explained before the new AAR dirty image improves the location estimation over the MVDR dirty image.
Furthermore, we go even beyond the location estimation, by improving the estimate of the sources intensity.
We suggest to estimate the intensities by using Least
Squares. Recall from (\ref{self_cal_model}) that
\[
\bea{lclcl}
\mR_k&=&\mA_k\mB\mA_k^H+\gs^2 \mI&=&
\sum_j\ba_k(s_j)\mI(s_j)\ba_k^H(s_j)+\gs^2 \mI\,.
\ena
\]
In each iteration we find a location $s_0$ that brings the MVDR
dirty image to its maximum at its direction. Thus, defining
$\ga=I(s_0)$, we solve the following problem:
\begin{equation} \label{mvdr_ls}
\bea{lcl}
    \hat{\ga}= \argmin_{\{\ga\}}
    \sum_{k=1}^K\;
    \|\; \mRh_{k} - \ga \ba_k(s_0)\ba_k^H(s_0)\|_F^2, && s.t.\hspace{ 2
    mm}
    \ga\geq0\,.
\ena
\end{equation}
where $\|\cdot\|_F$ is the Frobenius norm. Problem ({\ref{mvdr_ls}}) can be reformulated as
follows:
\begin{equation} \label{mvdr_ls1}
   \hat{\ga}= \argmin_{\{\ga\}} \|\vx-\ga\vh\|_2^2\,,
   \indent s.t. \hspace{2 mm} \ga\geq0\,,
\end{equation}
where
\[
  \bx=[vec(\mRh^{(1)})^T \hdots vec(\mRh^{(K)})^T ]^T
\]
is a vector of size $(Kp^2$x$1)$, in which $K$ is number of time
instances and $p$ is the number of antennas, the index $^{(k)}$
indicates the specific time instance, and
\[
    \vh=\left[\left(\ba^{(1)}(s_0)\otimes \bar{\ba}^{(1)}(s_0)\right)^T, \ldots,
    \left(\ba^{(K)}(s_0)\otimes \bar{\ba}^{(K)}(s_0) \right)^T\right]^T\,.
\]
The problem can be solved using the
Karush-Kuhn-Tucker Conditions for constrained optimization
\cite{boyd2004}. The result is
\begin{equation} \label{ls_solution}
    \hat{\ga}=\max\left \{\frac{\vh^H\bx}{\vh^H\vh},0 \right\}\,.
\end{equation}
When positivity constraint is not applicable, such as in polarimetric imaging, one similarly obtains
$\hat{\ga}=\frac{\vh^H\bx}{\vh^H\vh}$.
The MVDR dirty image $I_D'(\vs)$ can be replaced with the AAR dirty
image (\ref{AAR_dirty}) yielding better performance. The superiority
of the LS-MVI over the CLEAN method is shown in simulated examples
in section \ref{sec:comparison}.
%
\begin{table} [htb]
\caption{The LS-MVI algorithm}
\centering
\begin{tabular}{|l|}
\hline
  $n=0$.\\
  Set $\mRh_k^{(0)}=\mRh_k$. \\
  Set the loop gain $\grg$ (typically $\grg=0.1$).\\
  Calculate $\mI_D''$ according to Eq. (\ref{AAR_dirty}).\\
  last iteration flag$=$false.\\
 While not last iteration flag \\
 \quad $\bsh^{(n)}$ = $\argmax_{\{\bs\}}\;(|\mI_D''|)$.\\
 \quad Calculate $\hat{\ga}_n$ according to Eqs. (\ref{ls_solution}) and (\ref{mvdr_ls1}). \\
 \quad Improve the estimate of $\hat{\ga}_n$ using (\ref{eq:spd}). \\
  \quad $\mRh_k^{(n+1)}=\mRh_k^{(n)}-\gamma \hat{\ga}_n \ba_k(\bsh^{(n)})\ba_k(\bsh^{(n)})^H$, \quad $k=1,...,K$. \\
 \quad Calculate $\mI_D''$ using $\mR_k^{(n+1)}$. \\
 \quad n=n+1. \\
 Compute last iteration flag (Typically using $\chi^2$ test on residual image,\\ or using MDL).\\
 End.\\
 $\mI =  \sum_n \gamma\; \hat{\ga}_n\;\mB_{synth}(l-l_n, m-m_n)$.\\
\hline
\end{tabular}
\label{table_ls-mvi}
\end{table}
A possible improvement of the method above is by using semi-definite constraints. We can add the following
constraints on $\ga$ in (\ref{mvdr_ls1}):
\beq
\label{eq:spd}
\mRh_k-\gs^2\mI-\ga \va_k(s) \va_k(s)^H \succeq \vzero \quad k=1,...,K
\eeq
where $\mA\preceq \mB$ means that $\mB-\mA$ is positive semi definite.
Since the solution for $(\ref{mvdr_ls1})$ provides an upper bound on $\ga$ and $0$ is a lower bound, a simple
bi-section can provide this optimal value.
The LS-MVI algorithm is described in Table \ref{table_ls-mvi}.
\subsection{Accelerating the algorithm}
The proposed algorithm has higher computational complexity compared
to the CLEAN since it cannot utilize the two-dimensional FFT
algorithm to generate the MVDR dirty image.
It should be noted that naive implementation of the
techniques is much more complex than the CLEAN approach. It is still an interesting research problem to
reduce the complexity of the proposed parametric approach.
There are however several techniques that can accelerate the LS-MVI algorithm.
We would like to mention two approaches. First a clever update of the MVDR dirty image is possible, with complexity $KN^2p^2$, where $p$ is the number of antennas in the array, $N^2$ is the image size and $K$ is the number of observation epochs, by using low rank updates.  Moreover, since
the MVDR estimator better suppresses the effect of other sources, we
can estimate simultaneously several point sources with directional
vectors $s_1,...,s_d$, where $d$ is a small number, by choosing the
$d$ strongest points in the image. This is similar in spirit to the
small cycle in \cite{clark80} and \cite{schwab84}. To that end
define
\beq \vh_i=
    \left[\left(\ba^{(1)}(s_i) \otimes \bar {\ba}^{(1)}(s_i)\right)^T, \ldots,
    \left(\ba^{(K)}(s_i) \otimes \bar {\ba}^{(K)}(s_i) \right)^T\right]^T.
\eeq
and let $\mH=\left[\vh_1,...,\vh_d\right]$.
We can estimate the powers of the sources by solving the problem
\begin{equation} \label{mvdr_lsd}
   \hat{\ga}= \argmin_{\{\vga\}} \|\bx-\mH\vga\|^2\,,
   \indent s.t. \hspace{2 mm} \vga_i\geq0\,, i=1,...,d
\end{equation}
where $\vga=\left[\ga_1,\ldots,\ga_d \right]$ are the sources
powers. Similarly to the single source case, the solution is given
by solving the LS problem with positivity constraints. This problem
is a special case of quadratic programming problems, and therefore
can be solved efficiently. In contrast to \cite{briggs95}, where all
the measurement constraints are put into a large constrained LS
problem, our problem has low dimensional positivity constraints.
The solution is given by
\[
\ga_i=\max\left\{\left(\left(\mH^H\mH \right)^{-1} \mH^H \vx\right)_i,0 \right\}.
\]
Therefore, we solve an unconstrained LS solution and set the negative terms to $0$.
Solving for $d$ sources simultaneously reduces the complexity by a factor of $d$.
In this case we can also impose the extra conditions
\beq
\mRh_k^{(m)}-\sum_{i=1}^d \ga_i \va_k(\vs_i)\va_k(\vs_i)^H \succeq \vzero, \quad  k=1,...,K.
\eeq
These constraints prevent the matrices $\mRh_k^{(m)} $ from becoming non positive semi-definite and prevent
overestimation of the power.
This problem is equivalent to a special case of convex programming called semi-definite programming in the
variables $\vga=[\ga_1,...,\ga_d],t$
\beq
\bea{lcl}
\min t & & \\
s.t. &  & \\
\left[
\bea{cc}
\mI & \left(\vx-\mH \va\right) \\
\left(\vx-\mH \vga\right)^T & t
\ena
\right] & \succeq & \vzero \\
\sum_{i=1}^d \vga_i \va_k(\vs_i)\va_k^H(\vs_i) &\preceq& \mR_k \quad k=1,...,K \\
0 \leq \ga_i & & \\
0 \leq t & &
\ena
\eeq
After $\vgah=\left[\gah_1,...,\gah_d\right]^T$ is estimated, the covariance matrices $\mRh_k$ are updated, by subtracting the contribution of the estimated sources from the covariance matrices (possibly using a predefined loop gain)
\beq
\mRh^{(n+1)}_k=\mRh^{(n)}_k-\grg \sum_{i=1}^d \gah_i \va_k(\vs_i) \va_k^H(\vs_i)
\eeq
Such problems can be solved very efficiently using convex optimization techniques \cite{boyd2004}, especially, when the number of sources
estimated simultaneously is small.
Finally, we mention that the MVDR dirty image can be interpolated in the intermediate steps, resulting in
significant reduction in the complexity of the deconvolution.
\subsection{Self calibration and robust MVDR with a moving array}
We now turn to the case where the array response is not completely known, but we have some statistical
knowledge of the error, e.g., we know the covariance matrix of the array response error at each epoch.
Typically this covariance will be time invariant or will have slow temporal variation.
In this case we extend the robust dirty image as described in \cite{leshem2004}, into the moving array case.
This generalization is new, and has not been previously dealt with in the signal processing literature.
Since the positive definite constraint on the residual covariance matrices is important in our application,
we decided to extend the robust Capon estimator of \cite{stoica03}.
To that end assume that at each epoch we have an uncertainty ellipsoid describing the uncertainty of the
array response (as well as unknown atmospheric attenuation).
This is described by
\beq
\left(\va_k(\vs)-\vab_k(\vs) \right)^H \mC_k \left(\va_k(\vs)-\vab_k(\vs) \right) \le 1
\eeq
where $\vab_k(\vs)$ is the nominal value of the array response towards the point $\vs$.
Generalizing our MVDR with moving array we would like to solve the following problem:
\beq
\label{robust_capon_moving}
\bea{ll}
\left[\grh,\vah_1,...,\vah_k \right]=\arg \max_{\gr,\va_1,...,\va_k} \gr & \\
\hbox{subject to} & \\
\mRh_k- \gs^2 \mI-\gr \va_k \va_k^H\succeq \vzero & k=1,...,K  \\
\left(\va_k(\vs)-\vab_k(\vs) \right)^H \mC_k \left(\va_k(\vs)-\vab_k(\vs) \right) \le 1 & k=1,...,K.
\ena
\eeq
Let $\gt=1/\gr$.
The problem (\ref{robust_capon_moving}) is equivalent to the following problem
\beq
\bea{ll}
\left[\hat \gt,\vah_1,...,\vah_k \right]=\arg \min_{\gt,\va_1,...,\va_k} \gt & \\
\hbox{subject to} & \\
\left[
\bea{cc}
\mRh_k- \gs^2 \mI &  \va_k \\
\va_k^H           & \gt
\ena
\right]\succeq \vzero & k=1,...,K \\
\left[
\bea{cc}
\mC_k & \left(\va_k(\vs)-\vab_k(\vs) \right) \\
\left(\va_k(\vs)-\vab_k(\vs) \right)^H          & 1 \ena
\right]\succeq \vzero & k=1,...,K. \ena \eeq
This problem is once again a semi-definite programming problem that can be solved
efficiently via interior point techniques \cite{boyd96}. We can now
replace the MVDR estimator by this robust version. Interestingly we
obtain estimates of the corrected array response $\vah(\vs_k)$.
Using the model we obtain for each $k$ \beq \va_k(\vs)=\mgG
\vab(\vs). \eeq Hence, the self-calibration coefficients can be
estimated using least squares fitting
\beq \hat \mgG_k=\arg
\min_{\grg_1,...,\grg_p} \sum_{\ell=1}^L
\|\vah_k(\vs_\ell)-\mgG\vab_k(\vs_\ell)\|^2
\eeq
where
$\mgG=\diag\{\grg_1,...,\grg_p\}$. Of course, when the
self-calibration parameters vary slowly we can combine the
estimation over multiple epochs.
This might prove instrumental in calibration of LOFAR type arrays, where the calibration coefficients
vary across the sky. Since the computational complexity of the
self calibration semi-definite programming is higher than that of
the MVDR dirty image, it is too complicated to solve this problem
for each pixel in the image. Hence it should be used similarly to
the external self calibration cycle, where this problem is solved
using a nominal source locations model. The advantage over ordinary
self calibration, is that beyond the re-evaluation of the
calibration parameters, we obtain better estimates of the sources
powers, without significant increase in the complexity.
Another interesting alternative proposed by the anonymous reviewer is to use the doubly constrained RCB which combines norm constraint like in the AAR dirty image with the robust Capon beamforming \cite{li2004}. The extension to the moving array is done similarly to the previous problems, and will be omitted.
\section{Statistical analysis of the 2-D MVDR DOA estimator with a moving array}
\label{sec:MVDR_analysis}
In this section we analyze the two dimensional DOA estimator
based on MVDR with a moving array. The main motivation for this analysis is a first step
in analyzing the LS-MVI, but the results have independent value for DOA estimation with a moving array.
Interestingly, this is not a straightforward extension of the
analysis of the 2-D MVDR estimator with a fixed array by Hawkes and Nehorai \cite{Nehorai98}.
The proof is given in an appendix.

Vaidyanathan and Buckley \cite{vaidyanathan95} studied the
statistical properties of 1-D MVDR estimator and a fixed array. They
have used a single correlation matrix, and the unknown parameter was
the i-th source's location, represented by the scalar $\theta_i$,
$i=1,...,d$. Hawkes and Nehorai \cite{Nehorai98} considered the case
of 2-D MVDR estimator and a fixed array, i.e. they have also used a
single correlation matrix, but the unknown location was represented
by the vector $\bf{\gth}_i$ of size ($2\times1$). Our study extends
these works to the case of a moving array. The location of the i-th
source is denoted by $\bs_i = (l_i,m_i)$, $i=1,...,d$, where $l_i$
and $m_i$ are the coordinates on a plane that is as an approximation
of a small region in the celestial sphere (As described in Section
\ref{sec:astron}). $\mR_k$ is the correlation matrix at epoch $k$
where $k=1,...,K$ and $\mRh_k$ is the sample covariance matrix at
epoch $k$. For simplicity we assume that for each $k$, $\mRh_k$ is
based on $N$ samples (independent of $k$). This is realistic in most
applications.

The 2-D MVDR spectral estimator is given by:
\begin{equation} \label{max_fs_K}
    f(\bs)=\sum_{k=1}^K\frac{1}{\va_k\rH(\vs) \mRh_k^{-1}
    \va_k(\vs)}\,.
\end{equation}
When $K=1$, the function degenerates to
\begin{equation}
    f(\vs)=\frac{1}{\va\rH(\vs) \mRh^{-1}
    \va(\vs)}\,.
\end{equation}
Since maximizing $\frac{1}{\va\rH(\vs) \mRh^{-1} \va(\vs)}$ is equivalent to minimizing $1/f(\vs)=\va\rH(\vs)
\mRh^{-1}\va(\vs)$, both \cite{vaidyanathan95}, \cite{Nehorai98}
minimized $1/f(\bs)$ for scalar and vector $\vs$,
respectively. However, this approach can not be used with $K$
covariance matrices, because of the structure of $f(\bs)$ in (\ref{max_fs_K}). In order to be able
to generalize the analysis from a single correlation matrix to $K$
correlation matrices, we have to directly maximize (\ref{max_fs_K}). This technique significantly
complicates the analysis.

The estimator based on $N$ samples per covariance matrix is denoted
by $(\bsh_N)_i = ((\lh_N)_i, (\mh_N)_i)$. The estimation error is
given by
\beq
\gD \vs_N=\bsh_N-\vs.
\eeq Let $\vst=\left(\tilde{l},\tilde{m} \right)=\lim_{N \rarrow
\infty} \vs_N$ be the limit of the sequence of estimates. Based on
general estimation theory the limit exists with probability one.
Similarly to \cite{vaidyanathan95}, we decompose the estimation error
as
\beq
\gD\vs_N=\gD \vst_N+\gD \vst,
\eeq
where $\gD \vst=\vst-\vs$
is the asymptotic bias and $\gD \vst_N=\vsh_N-\vst$ is the finite
sample error. Note that $\gD \vst$ is a deterministic constant,
while $\gD \vst_N$ is a random variable. The variance of the finite
sample error (as well as the variance of the estimator) decays as
$O\left(\frac{1}{N}\right)$ and $\gD \vst_N$ converges to 0 with
probability 1. Explicitly
\[
    Var[(\lh_N)_i]=Var[(\mh_N)_i]=O(\frac{1}{N})\,.
\]
This implies that the variance of the 2D MVDR estimator converges to
zero asymptotically. The asymptotic bias can be explicitly computed. It is given by the following theorem:
%
%
\bthm  \label{asymptotic_bias1}

Let $\gD\bst_i =(\gD\lt_i,\gD\mt_i)$ be the asymptotic bias of the
MVDR estimator of the i-th source's location, $i=1,...,d$.

Then,
\begin{equation} \label{delta_lm}
   \gD\lt_i\simeq\frac{|A_1|}{|A|} \ \hbox{and} \ \gD\mt_i\simeq\frac{|A_2|}{|A|}\,,
\end{equation}
where
\[ |A| = \left| \begin{array}{cc}
\frac{\partial^2 f(\bs_i, \{\mR_k^{-1}\}_{k=1}^K)}{\partial l^2} & \frac{\partial^2 f(\bs_i, \{\mR_k^{-1}\}_{k=1}^K)}{\partial l\partial m} \\
\frac{\partial^2 f(\bs_i, \{\mR_k^{-1}\}_{k=1}^K)}{\partial l\partial
m} & \frac{\partial^2 f(\bs_i, \{\mR_k^{-1}\}_{k=1}^K)}{\partial m^2}
\end{array} \right|\]

\[ |A_1| = \left| \begin{array}{cc}
-\frac{\partial f(\bs_i, \{\mR_k^{-1}\}_{k=1}^K)}{\partial l} & \frac{\partial^2 f(\bs_i, \{\mR_k^{-1}\}_{k=1}^K)}{\partial l\partial m} \\
-\frac{\partial f(\bs_i, \{\mR_k^{-1}\}_{k=1}^K)}{\partial m} &
\frac{\partial^2 f(\bs_i, \{\mR_k^{-1}\}_{k=1}^K)}{\partial m^2}
\end{array} \right|\]

\[ |A_2| = \left| \begin{array}{cc}
\frac{\partial^2 f(\bs_i, \{\mR_k^{-1}\}_{k=1}^K)}{\partial l^2} & -\frac{\partial f(\bs_i, \{\mR_k^{-1}\}_{k=1}^K)}{\partial l} \\
\frac{\partial^2 f(\bs_i, \{\mR_k^{-1}\}_{k=1}^K)}{\partial l\partial
m} & -\frac{\partial f(\bs_i, \{\mR_k^{-1}\}_{k=1}^K)}{\partial m}
\end{array} \right|\,,\]

\begin{equation} \label{dfdlk}
 \frac{\partial f(\bs_i, \{\mR_k^{-1}\}_{k=1}^K)}{\partial l}= \sum_{k=1}^K \frac{4\pi Im(\mM_{k,2})}{(\mM_{k,1})^2}\,,
\end{equation}
\begin{equation} \label{dfdmk}
    \frac{\partial f(\bs_i, \{\mR_k^{-1}\}_{k=1}^K)}{\partial m}=
    \sum_{k=1}^K\frac{4\pi Im(\mM_{k,3})}{(\mM_{k,1})^2}\,,
\end{equation}
\begin{equation} \label{d2fdl2k}
\bea{ll}
    \frac{\partial^2 f(\bs_i, \{\mR_k^{-1}\}_{k=1}^K)}{\partial
    l^2}= \sum_{k=1}^K\frac{8\pi}{(\mM_{k,1})^3}[2Im^2(\mM_{k,2})
    \\ \hspace{30 mm}-\pi\mM_{k,1} Re(\mM_{k,4}-\mM_{k,5})]\,,
\ena
\end{equation}
\begin{equation} \label{d2fdm2k}
\bea{ll}
    \frac{\partial^2 f(\bs_i, \{\mR_k^{-1}\}_{k=1}^K)}{\partial
    m^2}=
    \sum_{k=1}^K\frac{8\pi}{(\mM_{k,1})^3}[2Im^2(\mM_{k,3})
    \\ \hspace{30 mm}-\pi\mM_{k,1} Re(\mM_{k,6}-\mM_{k,7})]\,, \ena
\end{equation}

\begin{equation} \label{d2fdldmk}
\bea{ll}
    \frac{\partial^2 f(\bs_i, \{\mR_k^{-1}\}_{k=1}^K)}{\partial l\partial
    m}=
    \sum_{k=1}^K\frac{8\pi^2}{(\mM_{k,1})^3}\{4Im(\mM_{k,2})Im(\mM_{k,3})
    \\ \hspace{30 mm}-\mM_{k,1} Re(\mM_{k,8}-\mM_{k,9})\}\,,
\ena
\end{equation}

\begin{equation} \label{M_definitions_k}
 \bea{ll}
    \mM_{k,1}=\ba_k^H\mRi_k\ba_k &  \mM_{k,6}=\ba_k^H\mV_k\mRi_k\mV_k\ba_k
\\
    \mM_{k,2}=\ba_k^H\mU_k\mRi_k\ba_k & \mM_{k,7}=\ba_k^H\mRi_k\mV_k\mV_k\ba_k
\\
    \mM_{k,3}=\ba_k^H\mV_k\mRi_k\ba_k & \mM_{k,8}=\ba_k^H\mU_k\mRi_k\mV_k\ba_k
\\
    \mM_{k,4}=\ba_k^H\mU_k\mRi_k\mU_k\ba_k & \mM_{k,9}=\ba_k^H\mU_k\mV_k\mRi_k\ba_k\,,
\\
    \mM_{k,5}=\ba_k^H\mRi_k\mU_k\mU_k\ba_k
 \ena
\end{equation}

\begin{equation}
    \ba_k=\ba_k(l,m)=\left[\bea{c} exp[-2\pi i(u_{11}(t_k)l+v_{11}(t_k)m)] \\ \vdots \\
exp[-2\pi i(u_{p1}(t_k)l+v_{p1}(t_k)m)]\ena\right]\,,
\end{equation}

\begin{equation}
    \mU_k =
    \left[ \bea{ccc}
    u_{11}(t_k) & & \mbox{\bf 0}\\
    & \ddots & \\
    \mbox{\bf 0} & & u_{p1}(t_k)
    \ena \right]
\end{equation}
and
\begin{equation}
 \mV_k =
    \left[ \bea{ccc}
    v_{11}(t_k) & & \mbox{\bf 0}\\
    & \ddots & \\
    \mbox{\bf 0} & & v_{p1}(t_k)
    \ena \right]\,.
\end{equation}

\ethm \vspace{3 mm}

The analytical expressions seem to be in a good agreement with
empirical values, as demonstrated in section
\ref{sec:simulated_bias}. The proof of the theorem is given in the appendix.

\section{Simulations}
\label{sec:simulation}
\begin{figure}
\centering
\subfigure[True image] 
{
    \label{fig:sub:a4}
    \includegraphics[width=5cm]{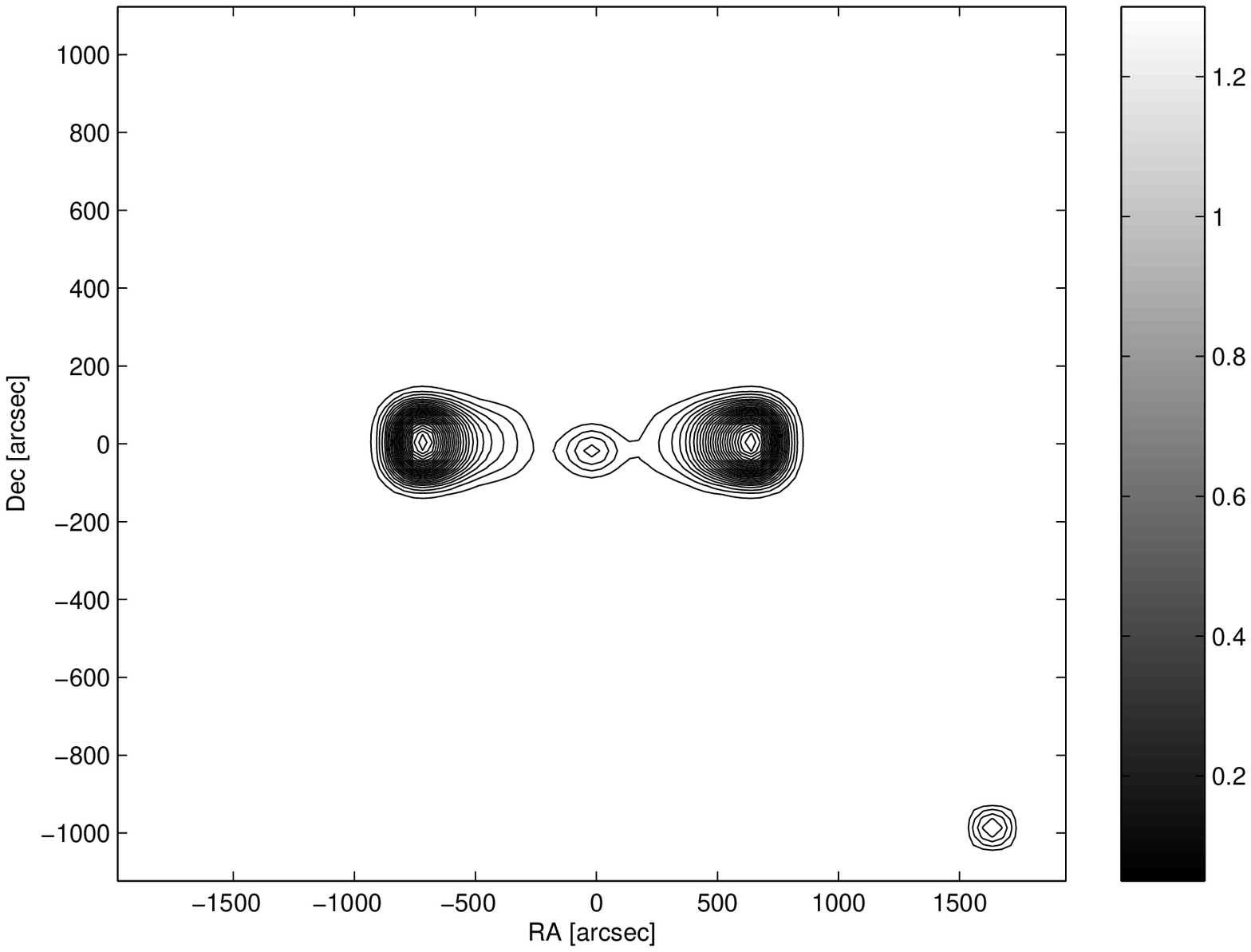}
} \hspace{1cm}
\subfigure[CLEAN image (100 iterations)]
{
    \label{fig:sub:b4}
    \includegraphics[width=5cm]{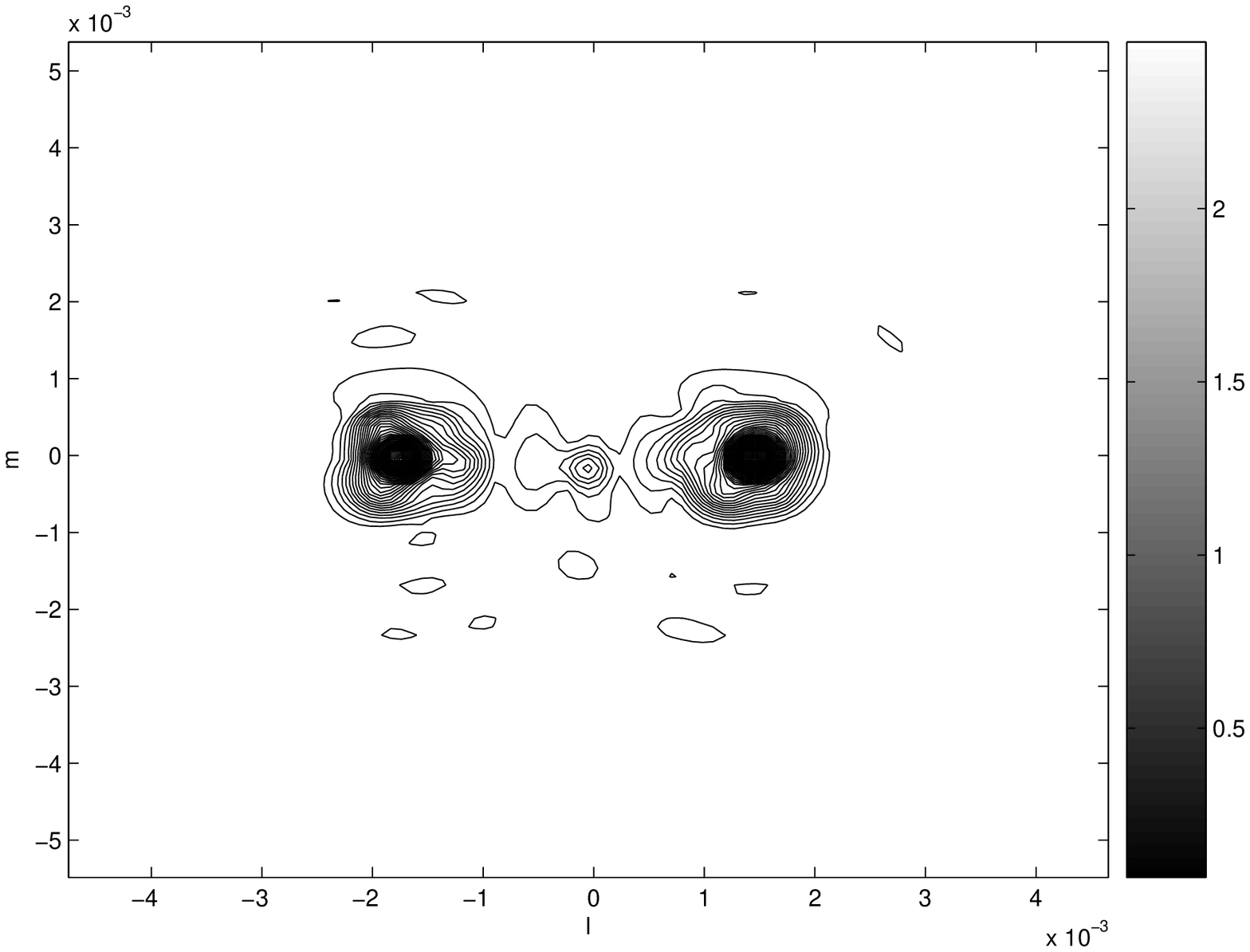}
} \\
\subfigure[CLEAN image (120 iterations)]
{
    \label{fig:sub:c4}
    \includegraphics[width=5cm]{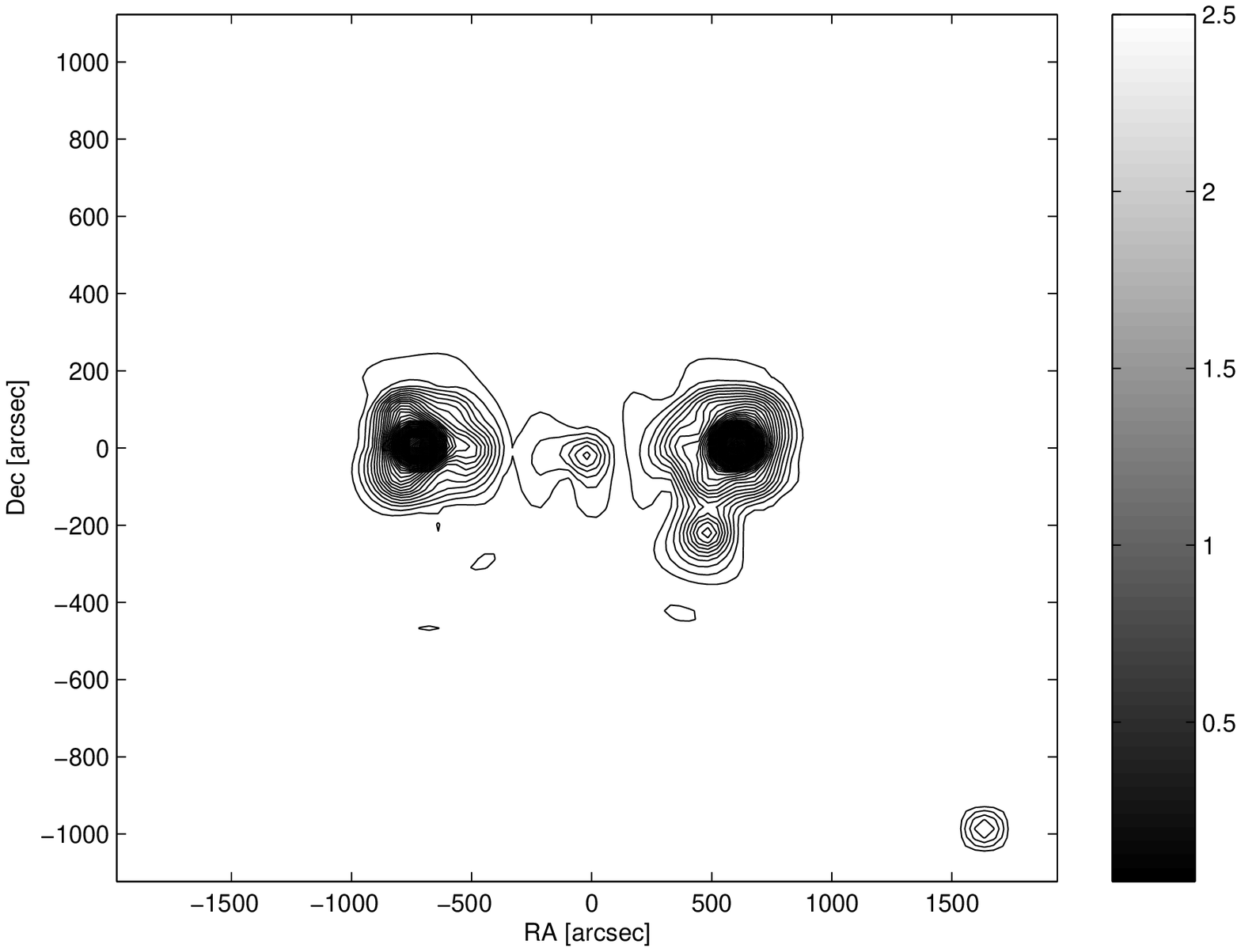}
}  \hspace{1cm}
\subfigure[LS-MVI image (100 iterations)]
{
    \label{fig:sub:d4}
    \includegraphics[width=5cm]{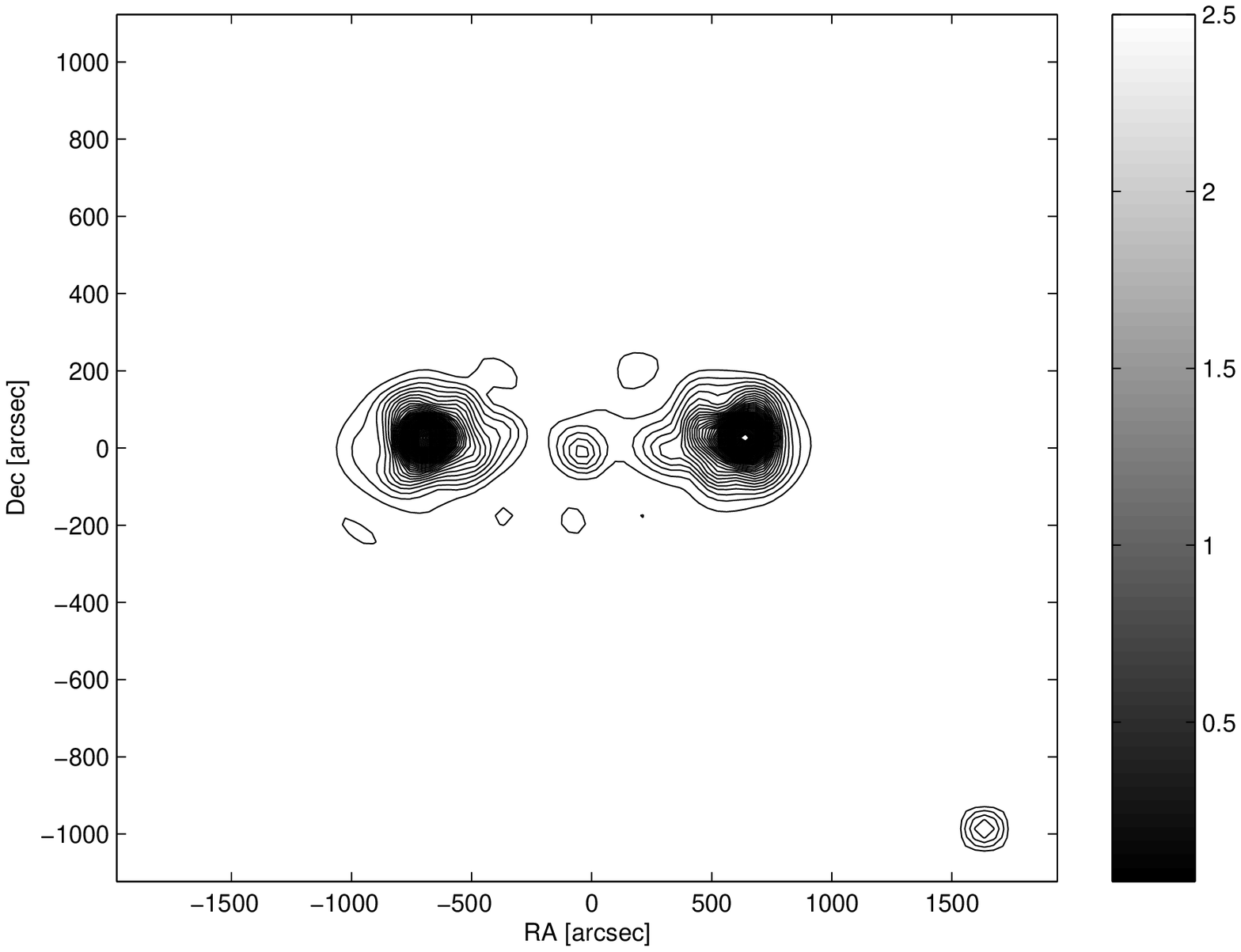}
}\\
\subfigure[LS-MVI image (300 iterations)]
{
    \label{fig:sub:e4}
    \includegraphics[width=5cm]{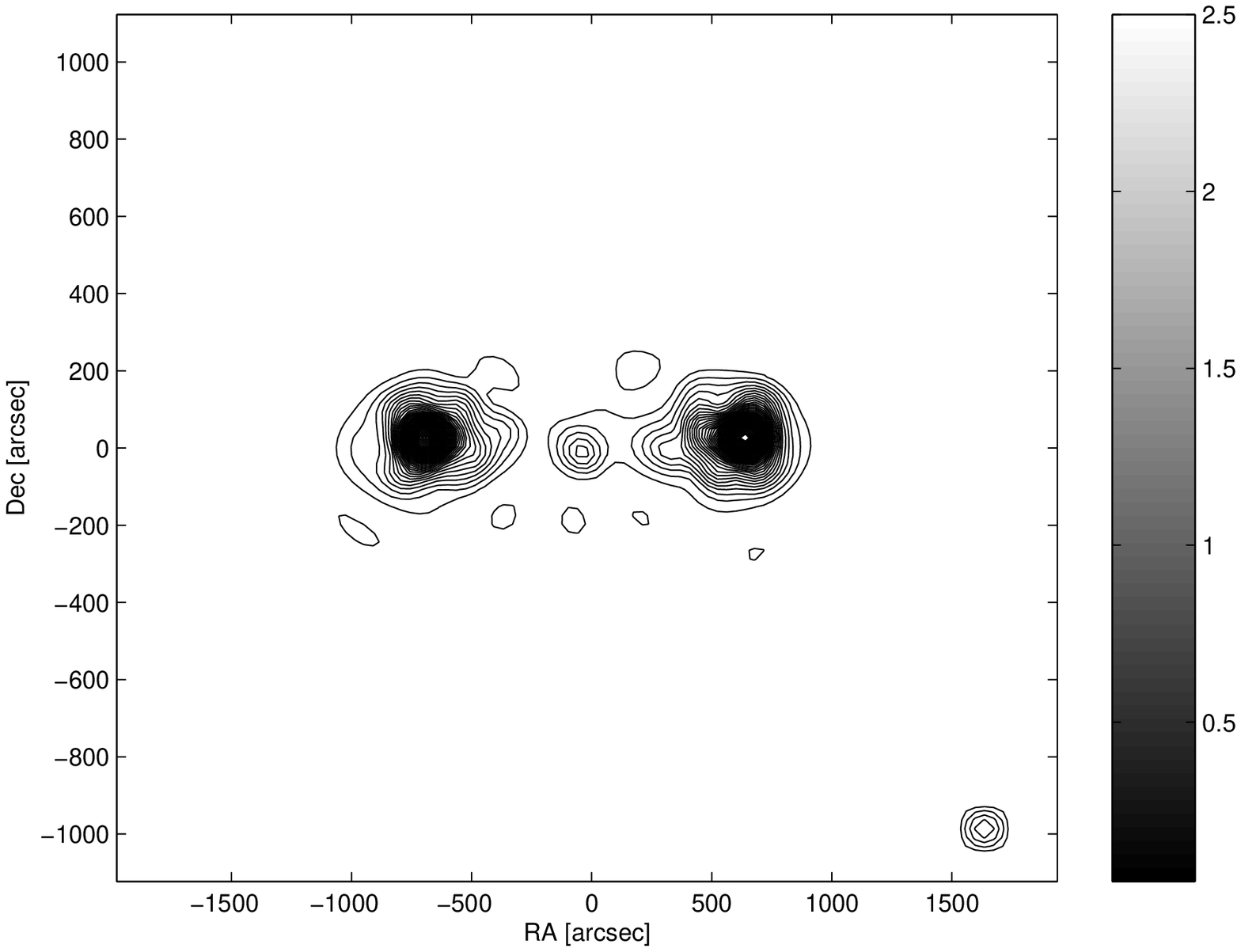}
}
\caption{Extended source. Power for weak source in the center is $3\gs$ per baseline. Contours of the image are $2\%$ of the maximum.
The bottom right is the synthesized beam with contours that are $20\%$ of the maximum of the synthesized beam.
            (a) The true image.
            (b) Image after CLEAN - 100 iterations.
            (b) Image after CLEAN - 120 iterations.
            (c) Image after  LS-MVI - 100 iterations.
            (c) Image after  LS-MVI - 300 iterations.
            }
\label{fig:expJ19} 
\end{figure}

In this section we present simulation results of the LS-MVI as well as of the asymptotic bias analysis.
In the first subsection we present deconvolution
results on simulated images. We compare the performance of the
LS-MVI algorithm and the CLEAN method. In the second subsection we study the asymptotic bias analysis.
In all simulations we have used an East-West
array of ten antenna elements, logarithmically spaced up to
$1000\lambda$ and generated artificial sky images. We converted the image to $(l,m)$
coordinates from right ascension and declination, as described in \cite{yen85}. Note that
this results in $(l,m)$ coordinates that are not on a rectangular grid. We produced 720 correlation
matrices along 12 hours, using the model of \cite{leshem2000b},
where the averaging period for each matrix was one minute. To simplify the simulations we assumed perfect coherence of the sources along each integration time and that compensation for the geometric delay  has been done.  The synthesized beam used in this paper is depicted in Figure \ref{fig:expJ19} at the right bottom
of each subfigure.
\subsection{Comparison of LS-MVI and CLEAN}
\label{sec:comparison}
Our first set of experiments compared the performance of the CLEAN algorithm and the LS-MVI.
A Gaussian noise was added to the measurements. To implement the CLEAN we created a uniform
grid and interpolated the visibilities to this grid using standard convolutional gridding. We created the
classical dirty image using fast Fourier inversion on the rectangular grid.

We applied the CLEAN algorithm on the created dirty image. Subsequently,
we created from the same data the MVDR dirty image (Def.
\ref{def:mvdr_dirty_image}) and the AAR dirty image (Def.
\ref{def:AAR_dirty_image}), based on the computed correlation
matrices. We performed deconvolution using the LS-MVI algorithm with the AAR dirty image, as
described in Table I. Both CLEAN and LS-MVI used a loop gain
$\gamma=0.1$ in all simulations. For each trial the original image,
the dirty image $I_D(s)$, the MVDR dirty image $I_{D}'(s)$ and the
AAR dirty image $I{''}_{D}(s)$ are presented. Then the results of
applying the CLEAN method and the LS-MVI algorithm. The latter was
based on the AAR dirty image.

\begin{figure}
\centering
\subfigure[True image] 
{
    \label{fig:sub:a2}
    \includegraphics[width=4cm]{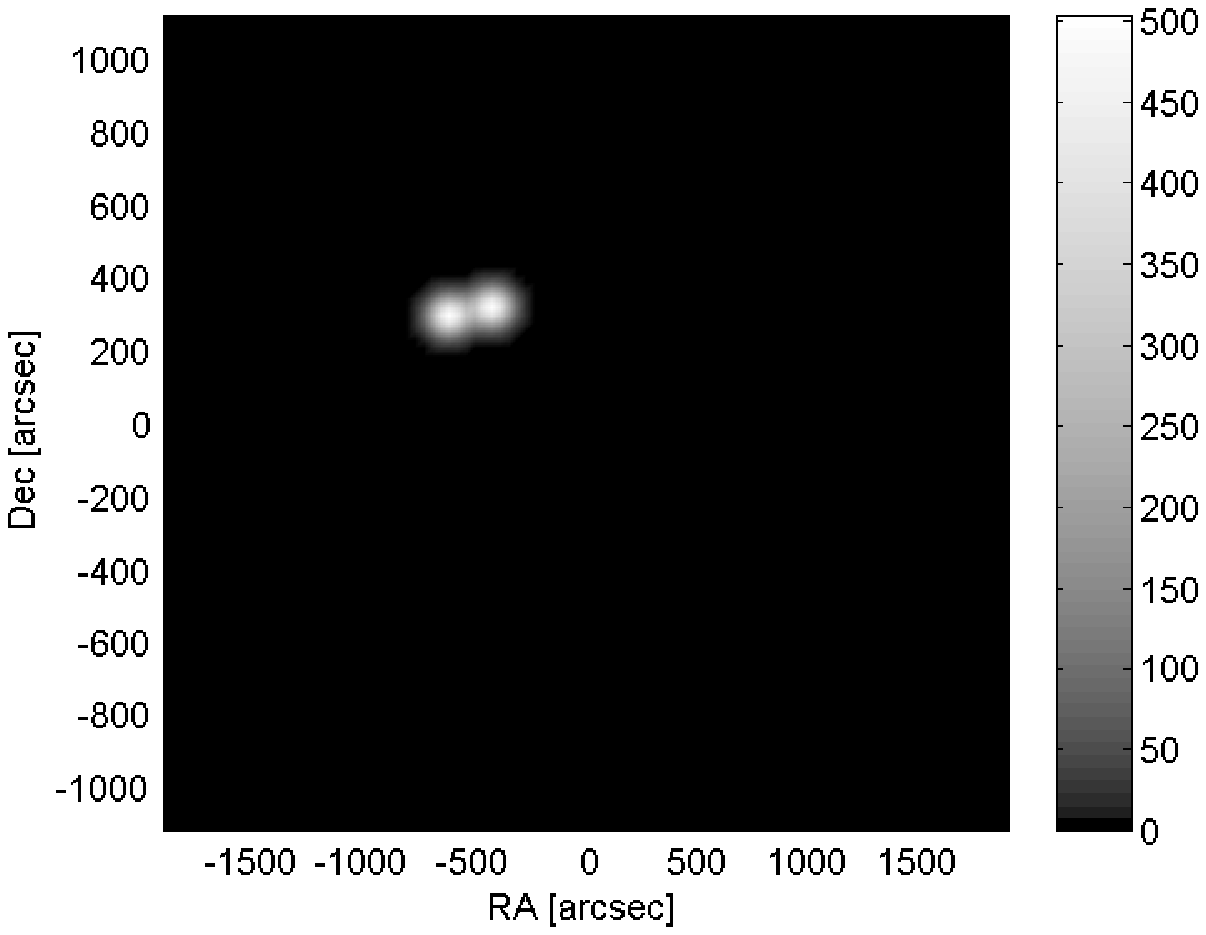}
} \hspace{1cm}
\subfigure[Dirty image] 
{
    \label{fig:sub:b2}
    \includegraphics[width=4cm]{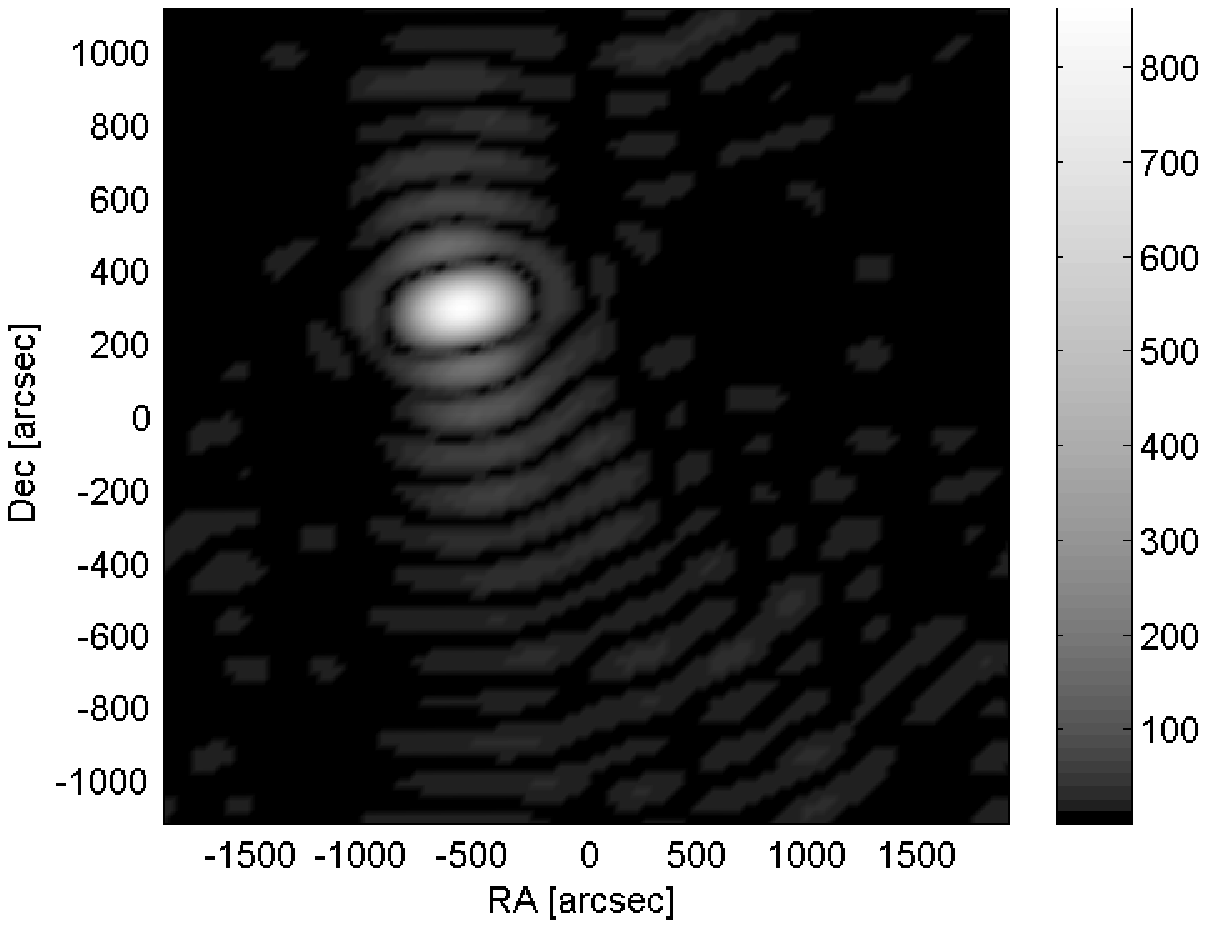}
} \hspace{1cm}
\subfigure[MVDR dirty image] 
{
    \label{fig:sub:c2}
    \includegraphics[width=4cm]{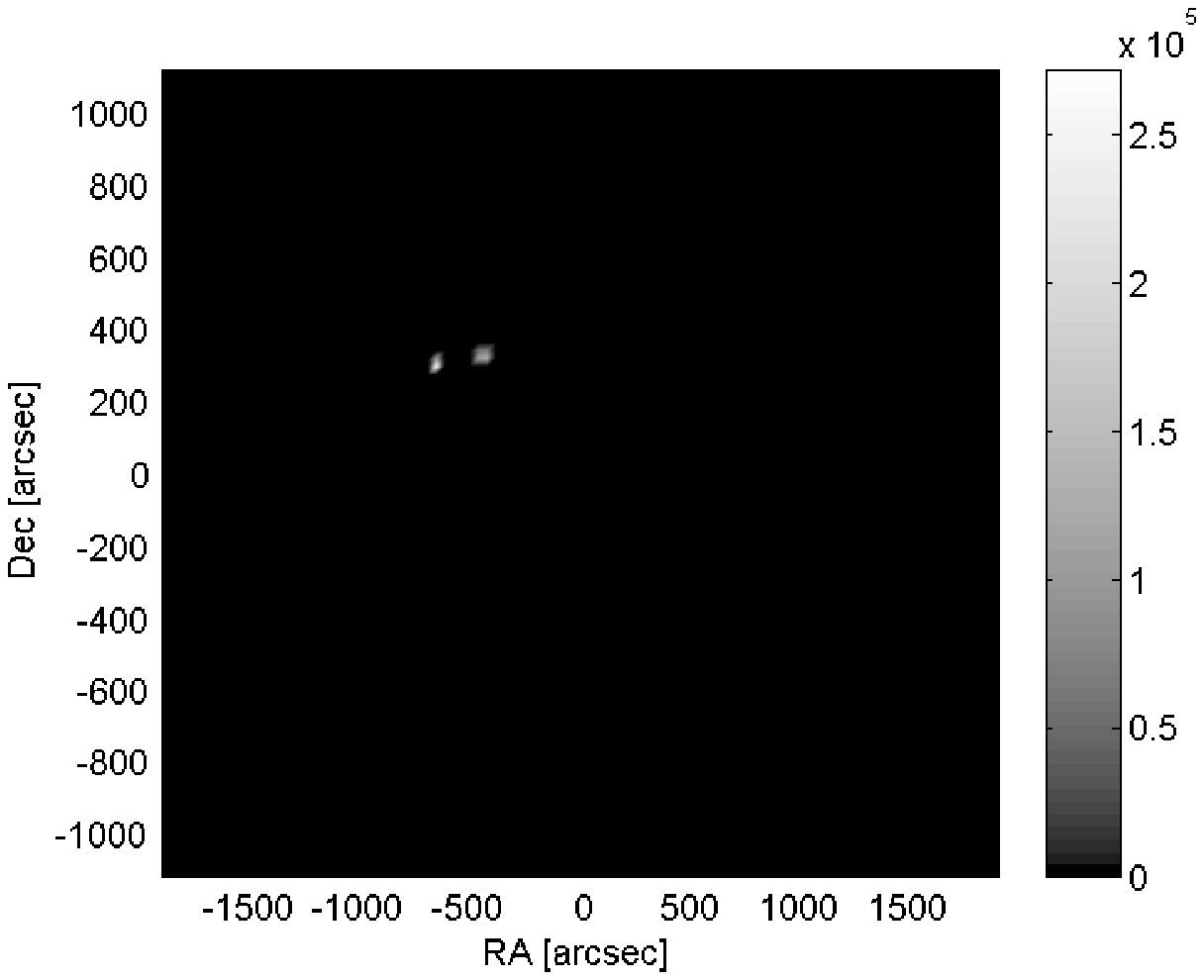}
} \\
\subfigure[AAR dirty image] 
{
    \label{fig:sub:d2}
    \includegraphics[width=4cm]{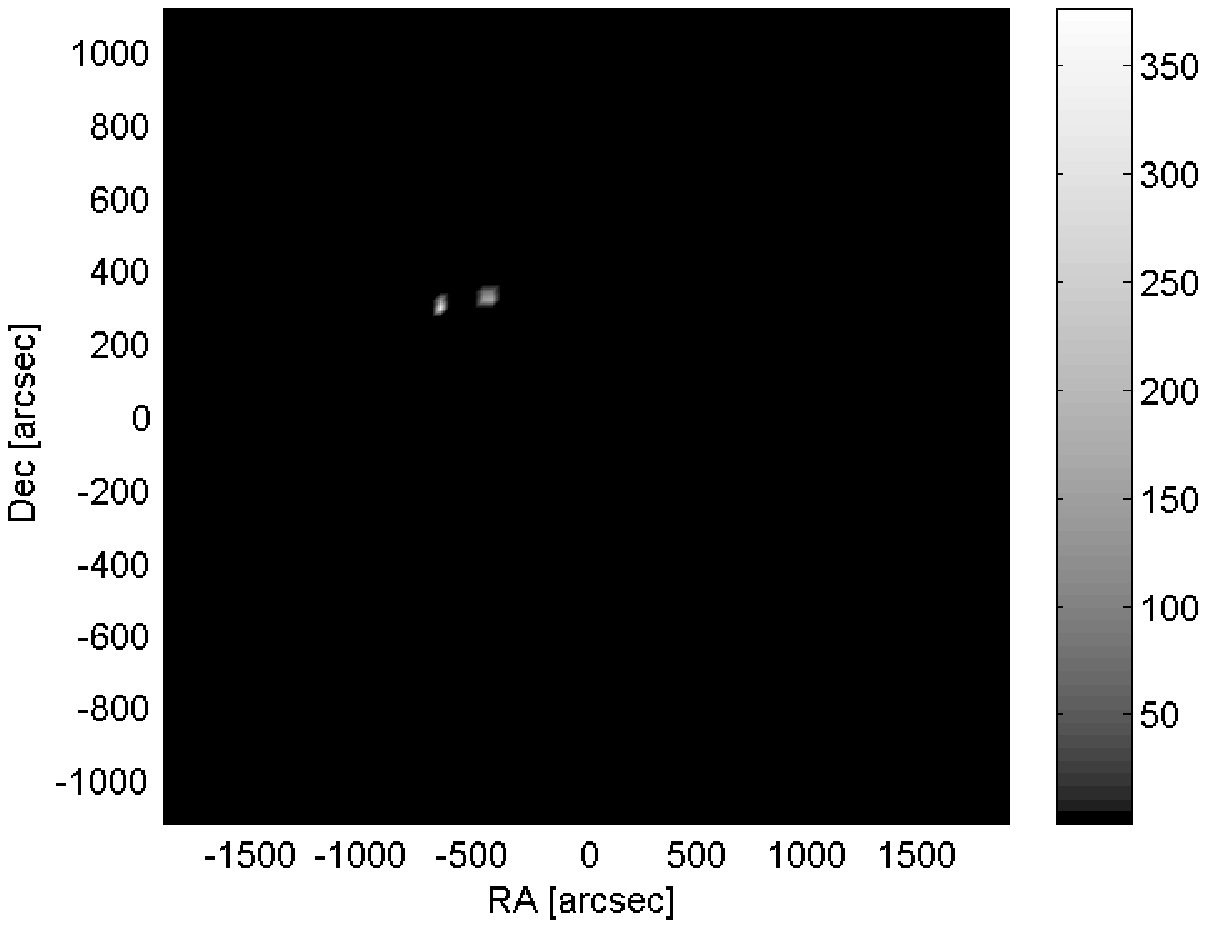}
}\hspace{1cm}
\subfigure[CLEAN image] 
{
    \label{fig:sub:d2}
    \includegraphics[width=4cm]{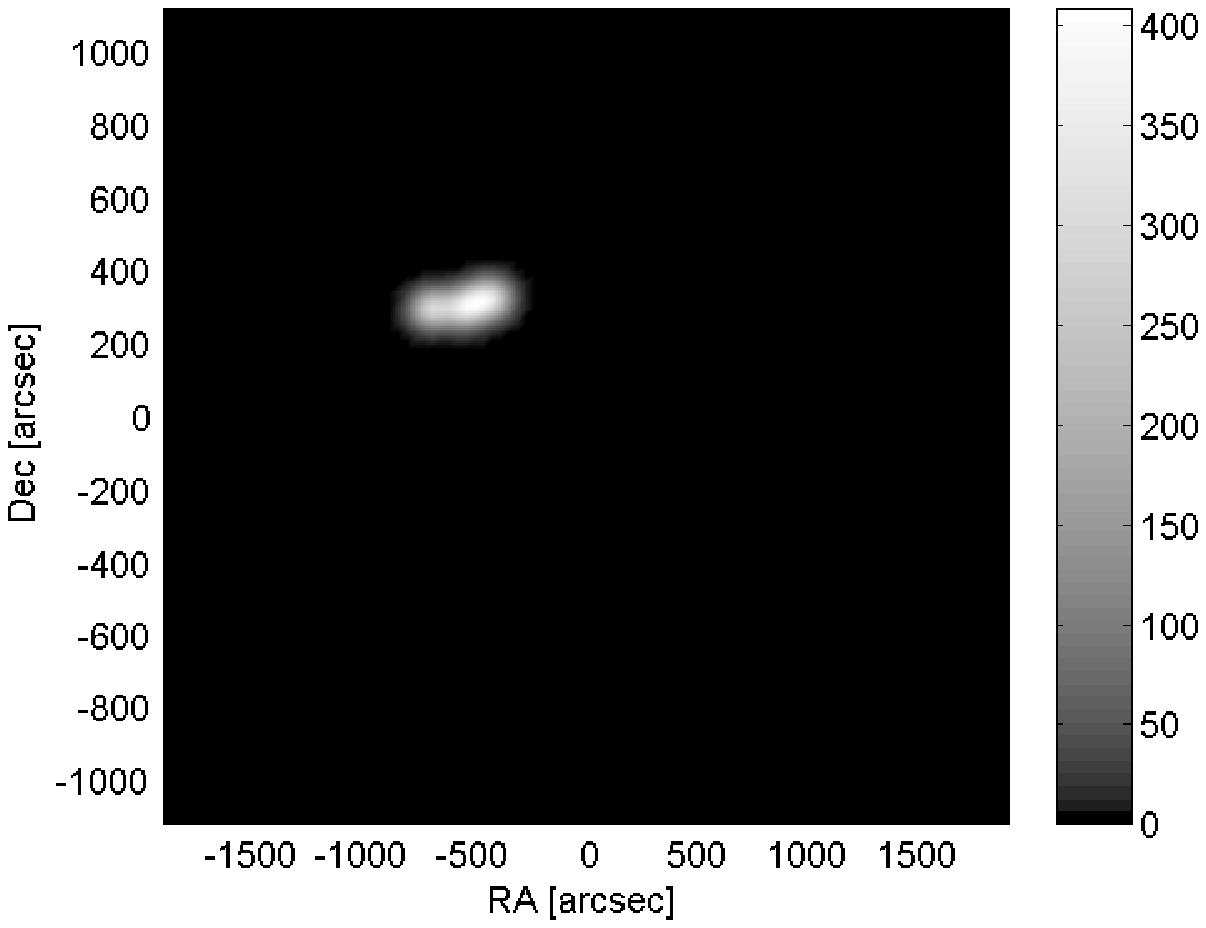}
} \hspace{1cm}
\subfigure[LS-MVI image] 
{
    \label{fig:sub:f2}
    \includegraphics[width=4cm]{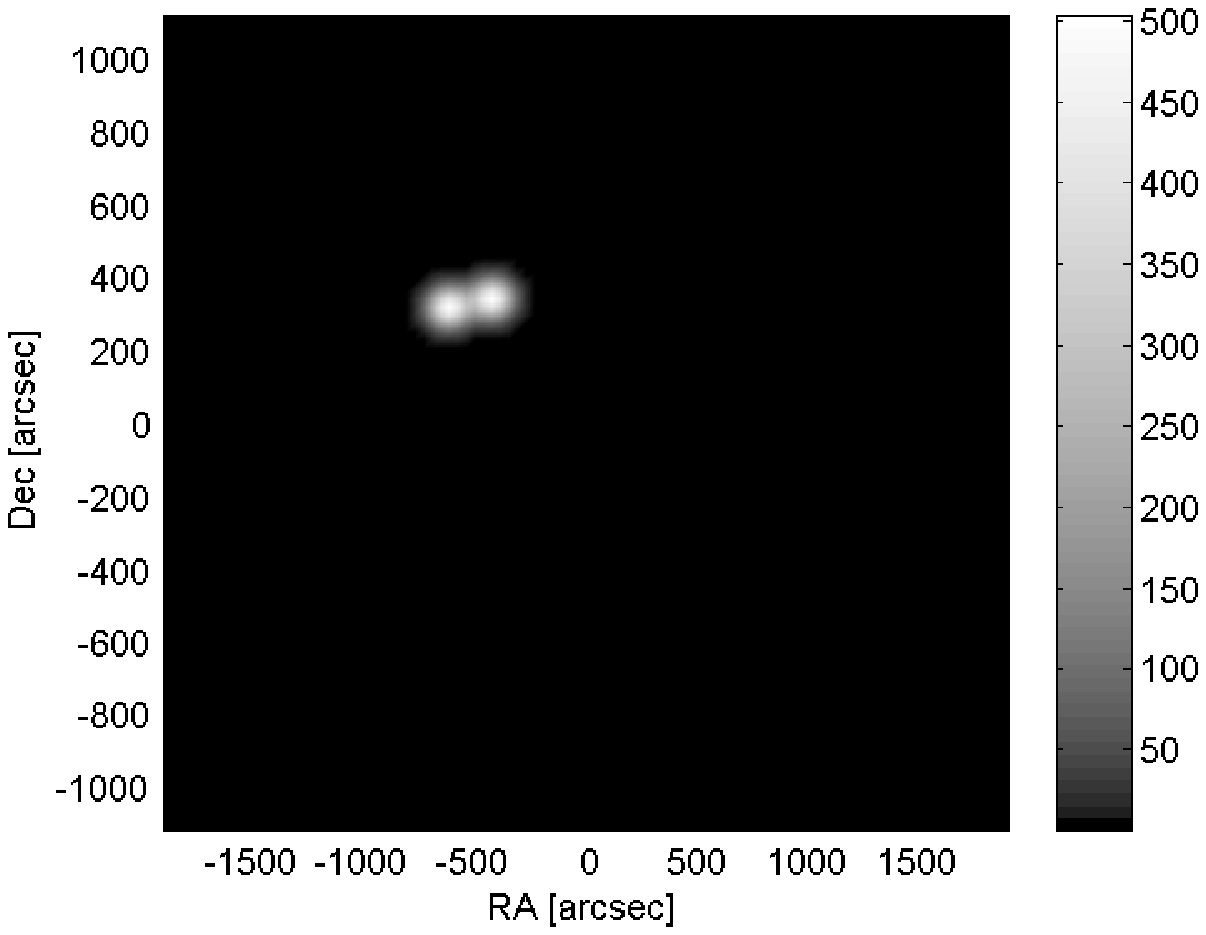}
}

\caption{Two very strong close sources. Signal power to noise standard deviation in the image was 250,000:1.
            (a) The true image.
            (b) The conventional dirty image.
            (c) The MVDR dirty image.
            (d) The AAR dirty image.
            (e) Image after CLEAN.
            (f) Image after  LS-MVI.}
\label{fig:exp2}
\end{figure}

The first scenario is depicted in Figure \ref{fig:exp2}. The
original image consists of two closely spaced sources. Their
intensities are equal and the noise was very weak, so that the image noise was 250,000 times weaker than each source. It is shown that LS-MVI
succeeds in separating the two sources whilst the CLEAN method does
not. Moreover, the intensities are more accurate in the LS-MVI's
image. Note that in this case the MVDR dirty image and the AAR dirty
image are quite similar.

\begin{figure}
\centering
\subfigure[True image] 
{
    \label{fig:sub:a3}
    \includegraphics[width=4cm]{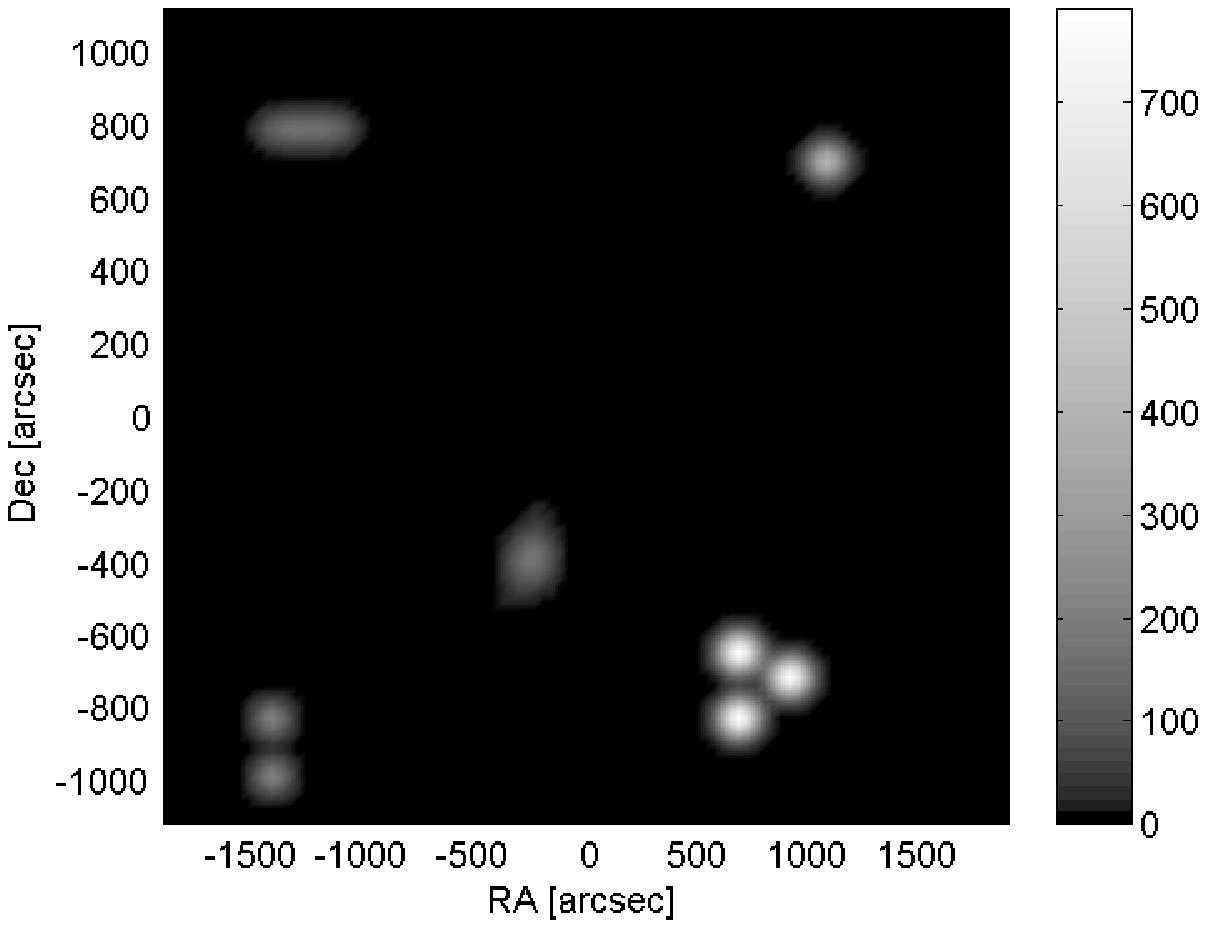}
} \hspace{1cm}
\subfigure[Dirty image]
{
    \label{fig:sub:b3}
    \includegraphics[width=4cm]{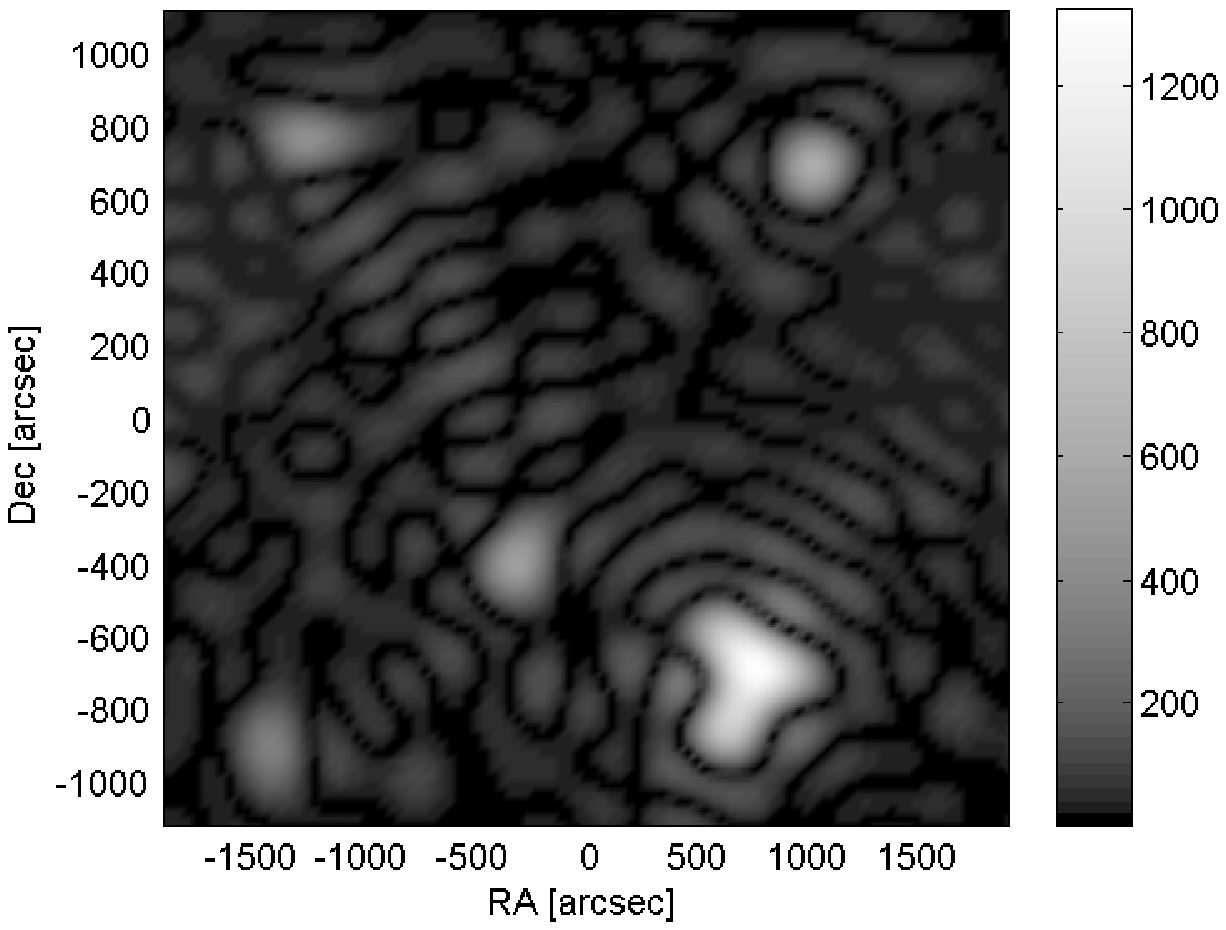}
} \hspace{1cm}
\subfigure[MVDR dirty image]
{
    \label{fig:sub:c3}
    \includegraphics[width=4cm]{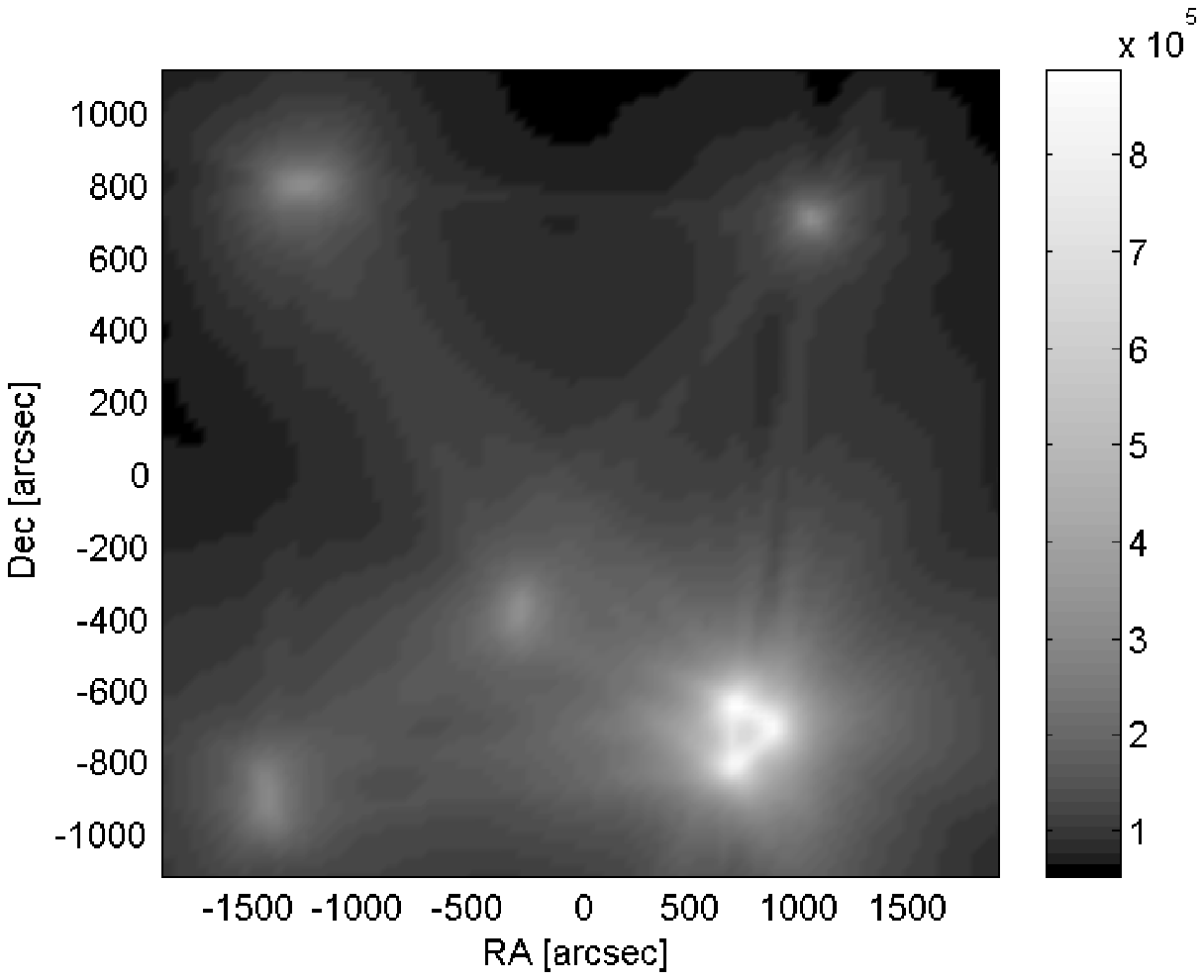}
}\\
\subfigure[AAR dirty image]
{
    \label{fig:sub:d3}
    \includegraphics[width=4cm]{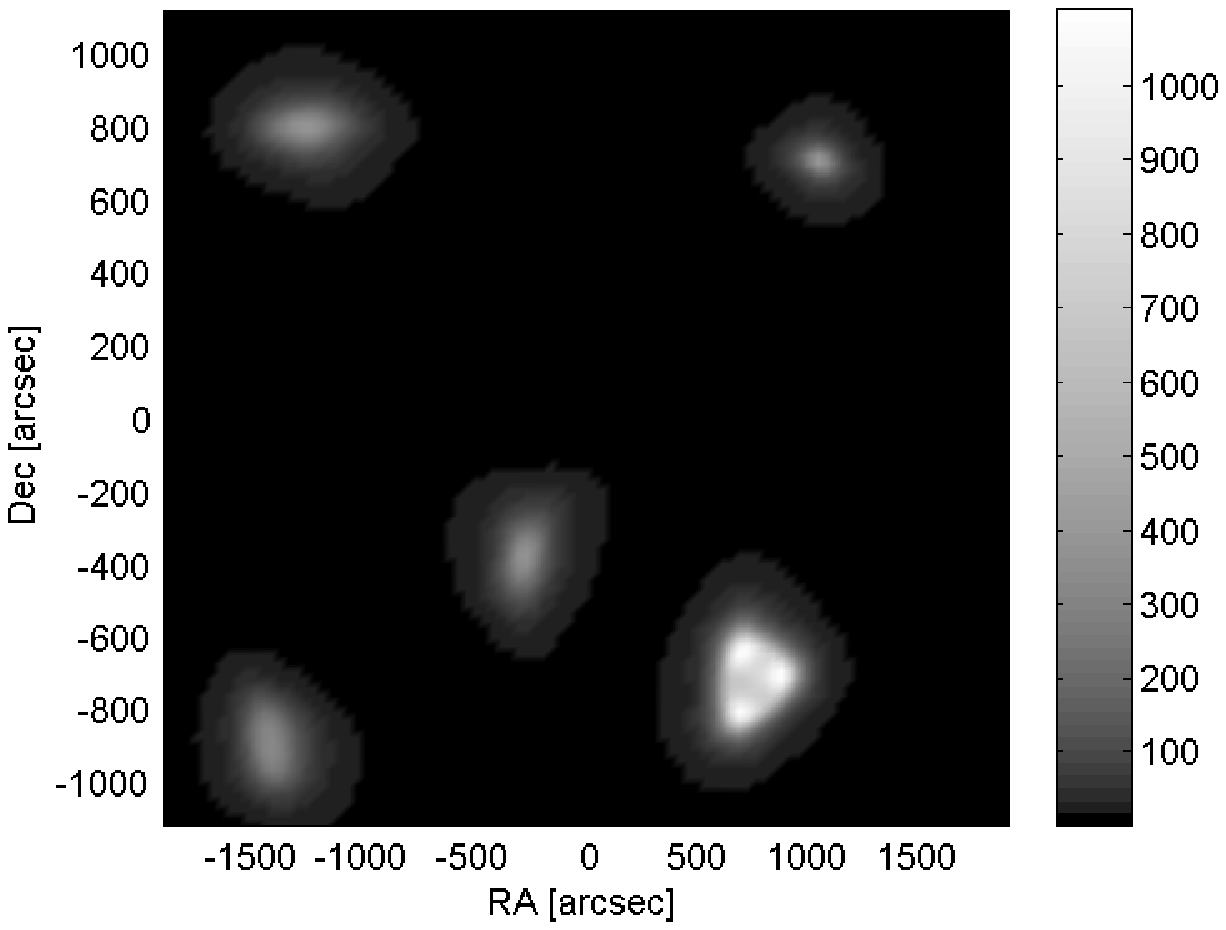}
} \hspace{1cm}
\subfigure[CLEAN image]
{
    \label{fig:sub:e3}
    \includegraphics[width=4cm]{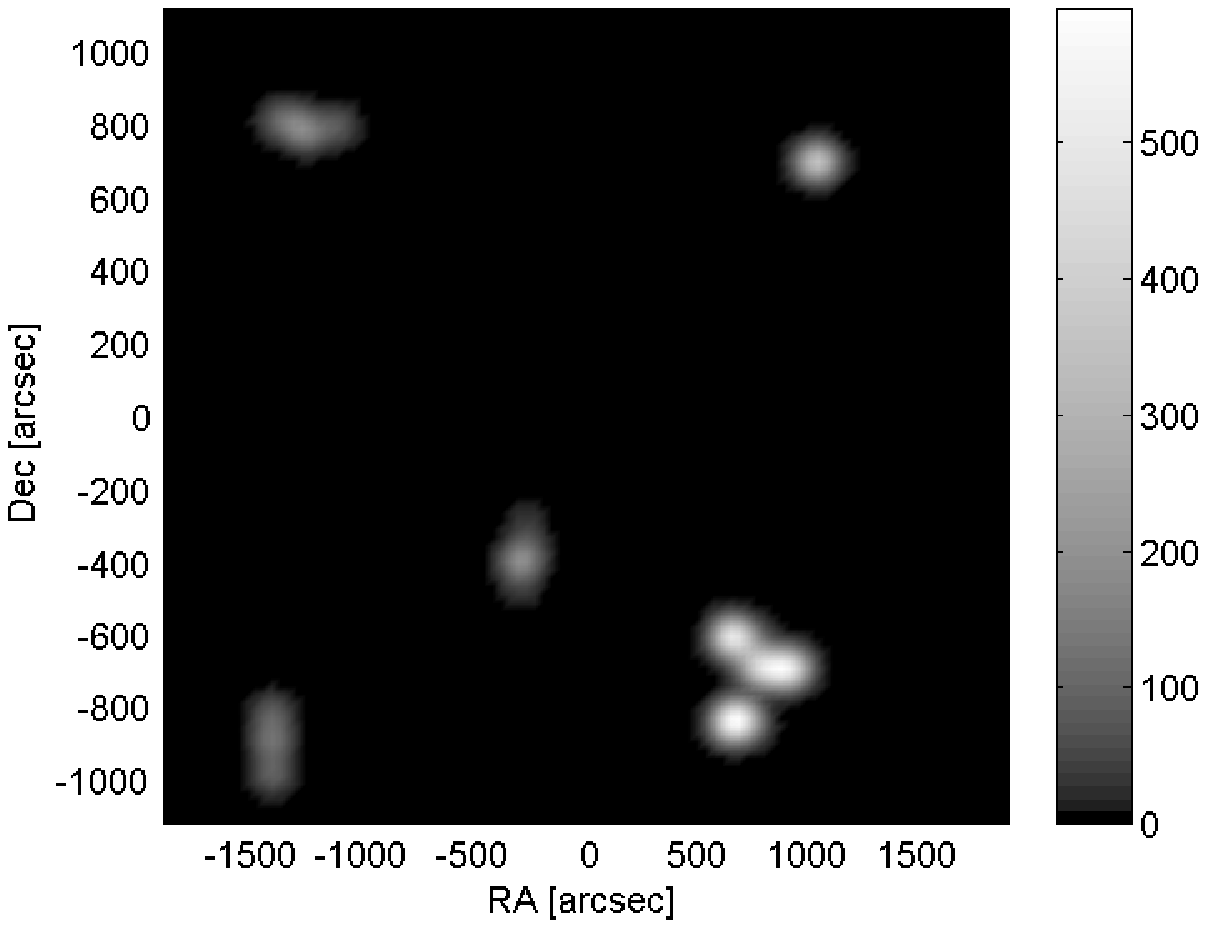}
} \hspace{1cm}
\subfigure[LS-MVI image]
{
    \label{fig:sub:f3}
    \includegraphics[width=4cm]{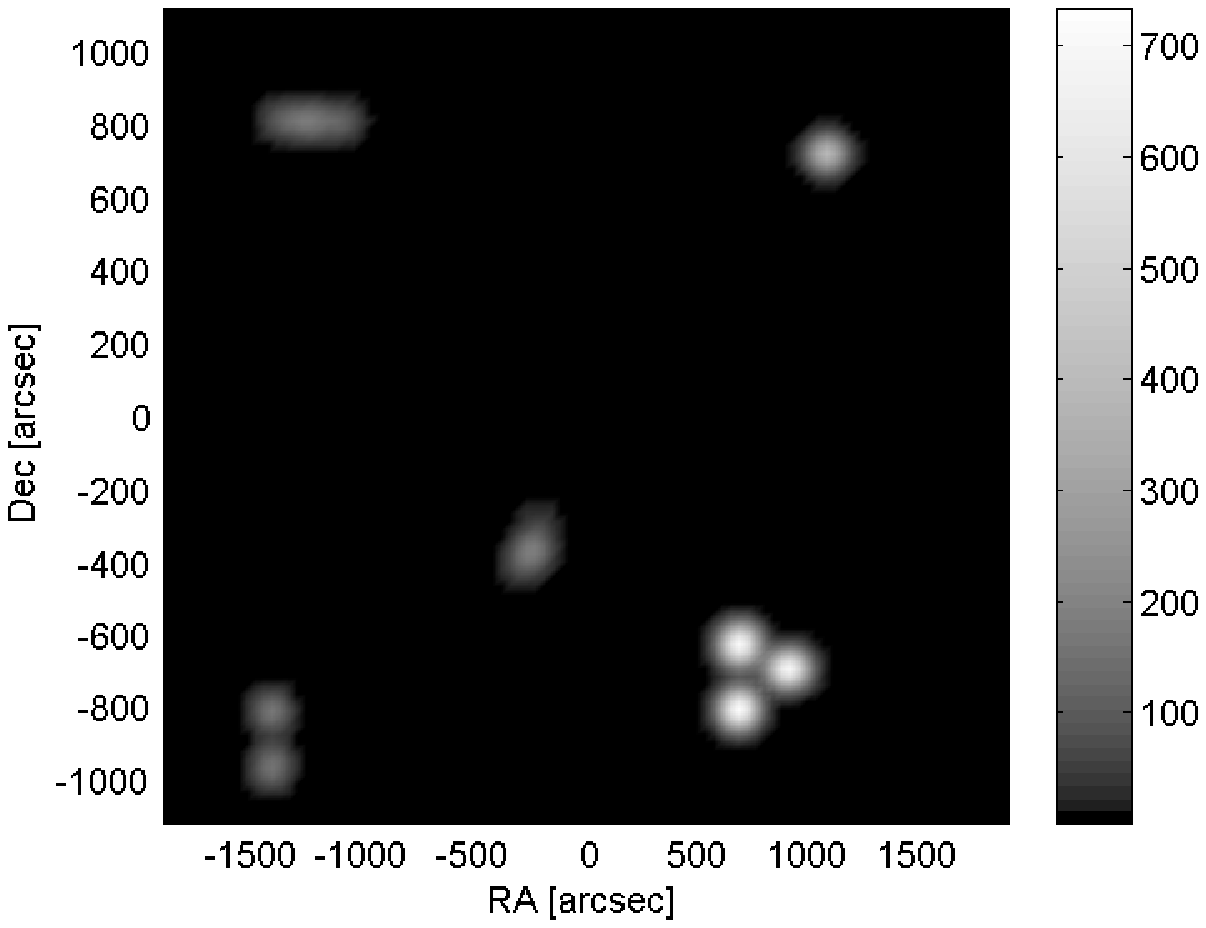}
}
\caption{Scattered sources. Intensities ratio between the strongest
source and the weakest source is 16db.
            (a) The true image.
            (b) The conventional dirty image.
            (c) The MVDR dirty image.
            (d) The AAR dirty image.
            (e) Image after CLEAN.
            (f) Image after  LS-MVI.}
\label{fig:exp3} 
\end{figure}

The second scenario is depicted in Figure \ref{fig:exp3}.
The original image consists of many sources with different intensities Fig:\ref{fig:sub:a3}. The intensities ratio
between the strongest source and the weakest source is 16dB. The weakest source contributed to each baseline
power that is equal to $1\gs$ of the noise standard deviation on the baseline. For the WSRT this is equivalent
to a source of 8 mJy when using 160 MHz band and integrating over 256 correlation lags \footnote{Based on the WSRT noise exposure calculator
http://www.astron.nl/~oosterlo/expCalc.html, and using the fact that the simulated beam is of size approximately 4 pixels}.
As can be seen, LS-MVI performs better than CLEAN. The latter cannot differentiate between close
sources, and the structures of its sources are less accurate.
Moreover, LS-MVI gives better intensities estimations than CLEAN. The reason for that is the isotropic noise spectrum,
that prevents direction dependent noise effects. Thus the sources
intensities estimations are more accurate when based on the AAR dirty image,
rather than on the conventional dirty image or on the MVDR dirty image.

\begin{figure}
    \begin{center}
%
    \includegraphics[width=0.35\textwidth]{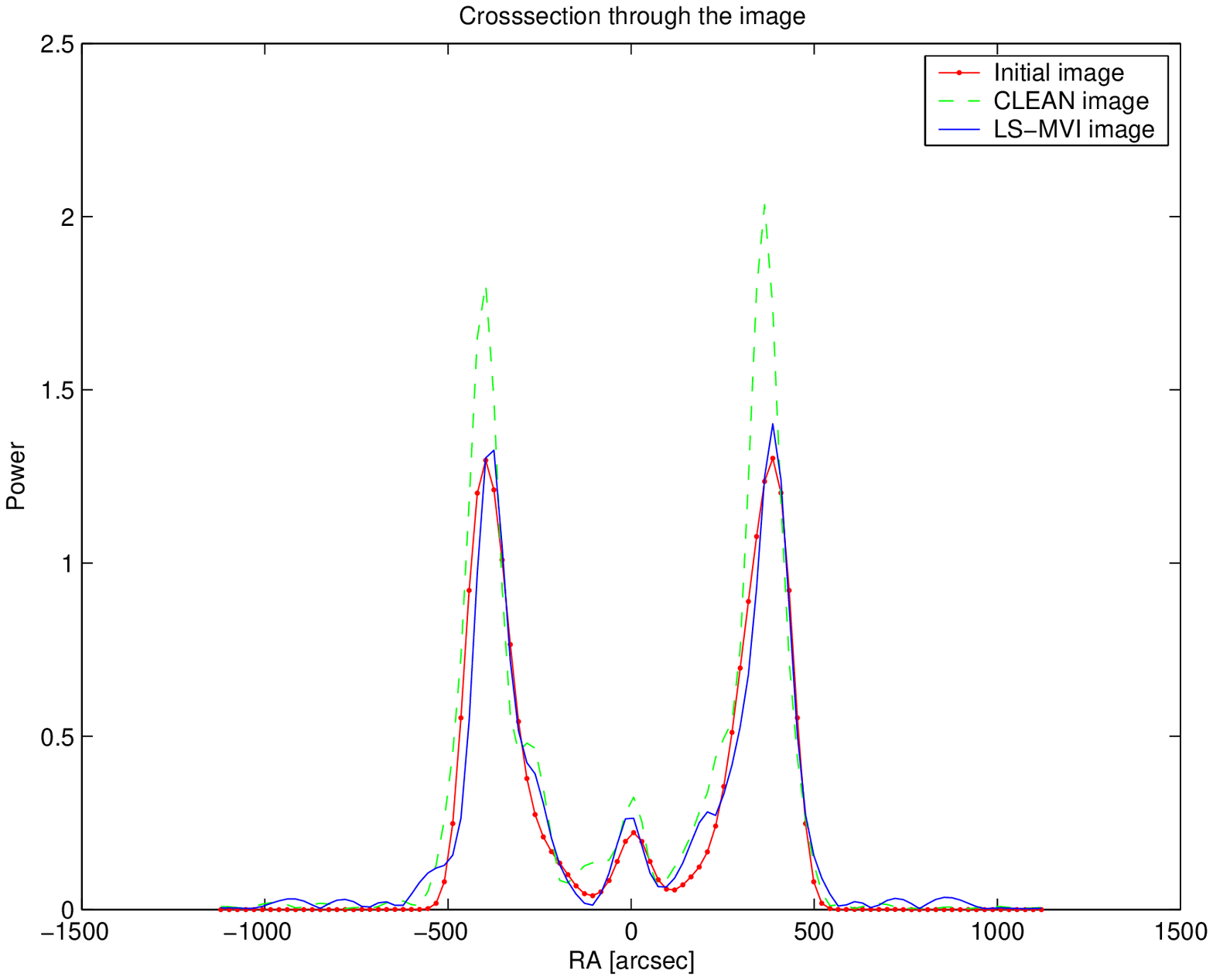}
    \end{center}
    \caption{Cross-section through the images.}
  \label{Fig:cross_section}
\end{figure}

The third experiment included an extended source containing a central weak point source and two
extended radio lobes. The correlation matrices were generated such that the contribution of the
weak source to each baseline was $3\gs$ of the noise on the baseline. The figure presents the original image,
the CLEAN image and the LS-MVI using the AAR dirty image. The CLEAN is presented after 100 and 120 iterations
while the LS-MVI is presented after 100 and 300 iterations respectively. Each contour is $2\%$ of the dynamic
range of the image.
We can clearly see that the CLEAN develops a fake source a little bit below the right lobe after 120 iteration.
This causes complete divergence if the CLEAN iterations are continued. On the other hand, the LS-MVI with its better power
estimate can continue with many more iterations, without affecting the image.
Finally to demonstrate the improved power estimate, we have taken a cross section of the images near the center.
Figure \ref{Fig:cross_section} presents the results, for the original image, the CLEAN after 100 iterations and the
LS-MVI after 100 iterations. We can clearly see the improved power estimate.

\subsection{Bias of the 2­D MVDR DOA estimator's simulation} \label{sec:simulated_bias}
\begin{figure}
    \begin{center}
    \includegraphics[width=0.35\textwidth]{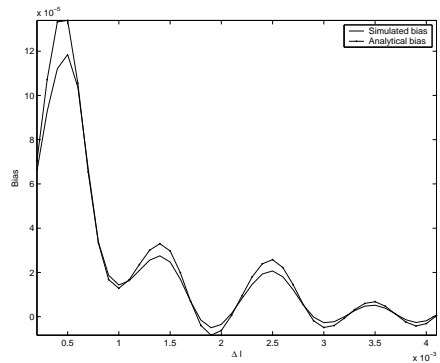}
    \end{center}
    \caption{Bias of the 2-D MVDR estimator}
  \label{Fig:bias}
\end{figure}

In this section we present a simulation of the bias of the 2­D MVDR
DOA estimator with a moving array. We have used the same array as in the previous simulations.
The artificial sky image was composed of two point sources with different intensities (the weak source had power that was half of the power of the strong source).
The sources contribution to each baseline was approximately 14$\sigma$ and 28$\sigma$ of the noise on the baseline.
The MVDR dirty image (Def. \ref{def:mvdr_dirty_image}) was created based on
these correlation matrices. First we found the location of the
maximum intensity in the MVDR dirty image. Then we used a 2-D
quadratic interpolation in order to obtain a fine 2-D MVDR location
estimate. We averaged the estimates over 100 independent trials.We compared our results to the
analytical expressions given in Theorem \ref{asymptotic_bias1}. This experiment was repeated 40 times
with various angular separation of the sources. Figure \ref{Fig:bias} depicts the simulated
bias ($\gD\lt$) against the analytical bias, as a function of the
angular distance between the two sources in l coordinates. It can be seen that the analytical results are in good
agreement with the simulation results. Moreover, as the distance
between the two sources increases, the bias of the estimator
decreases, as expected. As the distance is larger, the influence of
the sidelobes of one source on the other is smaller. The change in the bias is not monotonic in the angular
separation, since the weak source moves through the sidelobes of the strong source.
\section{Conclusions}
In this paper we extend the matrix formulation of \cite{leshem2000a} to non co-planar arrays and
polarimetric measurements. Then we propose a new parametric imaging technique that improves the resolution
and sensitivity over the classical CLEAN algorithm. The method is based on several improvements:
A new type of dirty image, LS estimation of the powers and semi-definite constraints.
We show how the technique can be combined into self-calibration using semi-definite programming.
Our semi-definite self-calibration algorithm also provides a new approach to robust beamforming with a
moving array, extending the techniques of \cite{li03}, \cite{vorobyov03} and \cite{lorenz05}.
We provide statistical analysis of the location estimator.
Simulated examples comparing full deconvolution using LS-MVI and comparing them to the
CLEAN method are presented. These simulations demonstrate that the parametric approach has higher resolution,
is more robust to source structures, and performs better in noisy situations.
The great potential of the methods proposed in this paper is a first step, towards the development of more
advanced imaging techniques, capable of providing higher dynamic range and interference immunity
as required by the radio telescopes of the future.

\section{Appendix}
In this section we derive approximate expressions for the asymptotic
bias by using a first-order Taylor series expansion of
$\frac{\partial f(\bst_i,\{\mR_k^{-1}\}_{k=1}^K)}{\partial l}$ and
$\frac{\partial f(\bst_i,\{\mR_k^{-1}\}_{k=1}^K)}{\partial m}$
around $\bs_i$. To reduce the notational load we present the
derivation of the single covariance matrix case. In the multiple
covariance matrices case $f$ is replaced by a sum and each
expression can be replaced by the appropriate sum over $k$. The
final results in theorem \ref{asymptotic_bias1} includes the
adjustment to multiple covariance matrices. Note that $\mR$ is a
fixed parameter representing the true covariance matrix, since we
discuss only the asymptotic term. Let $\mR^{-1}$ be fixed. The
expansions around $\bs$ is given by:
\begin{equation} \label{taylorl_r1}
\bea{lcl}
   \frac{\partial f(\bst_i, \mR^{-1})}{\partial l}
=    \dfdl+\dfTdlT\cdot\gD l_i
\\
    \hspace{3 mm} +\dfTdmdl\cdot\gD m_i+o(|(\gD l_i|)+o(|\gD
   m_i|)
\ena
\end{equation}
and
\begin{equation}  \label{taylorm_r1}
\bea{lcl}
   \frac{\partial f(\bst_i, \mR^{-1})}{\partial m}
   =\dfdm+\dfTdmT\cdot\gD m_i
\\
    \hspace{3 mm}  +\dfTdldm\cdot\gD l_i+o(|(\gD l_i|)+o(|\gD
    m_i|)\,.
\ena
\end{equation}
Since by definition $\vst$ is the minimum of $f(\bst_i, \mR^{-1})$
\[
\bea{lcl}
 \frac{\partial f(\bst_i, \mR^{-1})}{\partial l}=0, &&
 \frac{\partial f(\bst_i, \mR^{-1})}{\partial m}=0.
\ena
\]
%
Hence,
\begin{equation}
 \bea{lcl}
    \dfTdlT \cdot \gD l_i + \dfTdmdl \cdot \gD m_i \simeq -\dfdl
    \\
    \dfTdldm \cdot \gD l_i + \dfTdmT \cdot \gD m_i \simeq -\dfdm\,.
\ena
\end{equation}
 Proving that $\dfTdmdl=\dfTdldm$ and using Cramer's rule, we
obtain Eqs. (\ref{delta_lm})  (for k=1).

\vspace{3 mm} For simplicity, from now on we will denote
$\ba(\bs_i)$ by $\ba$.
\newline Let us now derive the expressions for the derivatives of
$f(\bs_i,\mR^{-1})$.
\begin{equation}
    f(\bs_i, \mR^{-1}) \;=\;
        \;\frac{1}{\ba^H\mR^{-1}\ba}
\end{equation}
Therefore,
\begin{equation} \label{derivative_l}
    \dfdl=\frac{-\frac{\partial}{\partial
    l}[\ba^H\mR^{-1}\ba]}{[\ba^H\mR^{-1}\ba]^2}\,,
\end{equation}
where
\begin{equation} \label{derivative_mul}
 \frac{\partial}{\partial
l}[\ba^H\mR^{-1}\ba]=\frac{\partial\ba^H}{\partial
l}\mR^{-1}\ba+\ba^H\mR^{-1}\frac{\partial \ba}{\partial l}\,.
\end{equation}
\vspace{5 mm} Using  the fact
$\mR^{-1}$ is Hermitian and simple algebraic manipulation yields
\[
    \frac{\partial\ba^H}{\partial
    l}\mR^{-1}\ba+\ba^H\mR^{-1}\frac{\partial \ba}{\partial
    l}=2Re(\ba^H\mR^{-1}\frac{\partial \ba}{\partial
    l})\,.
\]
Therefore,
\begin{equation} \label{derivative_mul_final}
    \frac{\partial}{\partial
    l}[\ba^H\mR^{-1}\ba]=2Re(\ba^H\mR^{-1}\frac{\partial \ba}{\partial
    l})\,.
\end{equation}
Using Eq. (\ref{derivative_mul_final}),
 Eq. (\ref{derivative_l}) becomes
\begin{equation} \label{dfdl_detailed}
    \dfdl=
  \frac{4\pi
    Im(\mM_2)}{(\mM_1)^2}\,,
\end{equation}
where we have used the fact that both $\mU$ and $\mRi$ are
Hermitian, $\mM_1$ and $\mM_2$ are defined in Eq.
(\ref{M_definitions_k}) for k=1. Thus we get Eq. (\ref{dfdlk}) (for
k=1). Similarly we derive Eq. (\ref{dfdmk}).

To calculate $\frac{\partial^2
f(\bs_i, \mR^{-1})}{\partial l^2}$ (Eq. (\ref{d2fdl2k})), define $g$  as
\begin{equation} \label{denote_g}
    g(\bs_i, \mR^{-1}) = 2Re(\ba^H\mR^{-1}\frac{\partial \ba}{\partial
    l})\,.
\end{equation}
Then,
\begin{equation} \label{denote_f}
  \frac{\partial f(\bs_i, \mR^{-1})}{\partial l}= -\frac{g(\bs_i, \mR^{-1})}{(\ba^H\mR^{-1}\ba)^2}
\end{equation}
and
\[
   \frac{\partial g(\bs_i, \mR^{-1})}{\partial
    l} = 2Re(\aRdaTdlT+\dadlRdadl) \,.
\]
Therefore,
\begin{equation} \label{d2fdl2_not_final}
\bea{lcl}
   \frac{\partial^2 f(\bs_i, \mR^{-1})}{\partial
    l^2}
    =
    \frac{2}{(\mul)^3}[4Re^2(\aRdadl)
\\ \hspace{3 mm}
    -(\mul)Re(\aRdaTdlT+\dadlRdadl)\,.
 \ena
\end{equation}
Using again the fact that both
$\mU$ and $\mRi$ are Hermitian, we get
\begin{equation}
\bea{lcl}
   \frac{\partial^2 f(\bs_i, \mR^{-1})}{\partial
    l^2}
    = \frac{8\pi}{(M_1)^3}[2Im^2(\mM_2)-\pi\mM_1
    Re(\mM_4-\mM_5)]\,,
 \ena
\end{equation}
where $\mM_1$, $\mM_2$, $\mM_4$ and $\mM_5$ are defined in Eq.
(\ref{M_definitions_k}) for k=1. Thus we get  Eq. (\ref{d2fdl2k})
for k=1. Eq. (\ref{d2fdm2k}) may be derived in the same way.

To calculate $\frac{\partial^2 f(\bs_i, \mR^{-1})}{\partial
l\partial m}$ (Eq. (\ref{d2fdldmk})). we use  Eq. (\ref{denote_g})
and obtain
\begin{equation} \label{dg_dm}
    \frac{\partial g(\bs_i, \mRi)}{\partial
    m} =2Re(\frac{\partial \ba^H}{\partial m}\mRi\frac{\partial \ba}{\partial l}+
    \ba^H\mRi\frac{\partial^2 \ba}{\partial l \partial m})\,.
\end{equation}
By (\ref{denote_f}) and (\ref{dg_dm}), we get
\begin{equation} \label{df2dldm_not_final}
\bea{lcl}
    \frac{\partial^2 f(\bs_i, \mR^{-1})}{\partial l\partial m} =
     \frac{2}{(\mul)^3}[4Re(\aRdadl)
     Re(\aRdadm)
\\
    \hspace{7 mm} -(\mul)Re(\frac{\partial \ba^H}{\partial m}\mRi\frac{\partial \ba}{\partial l}+
    \ba^H\mRi\frac{\partial^2 \ba}{\partial l \partial m})]\,.
 \ena
\end{equation}

After some simple algebraic manipulations we get Eq.
(\ref{d2fdldmk}) for k=1. Note that $\frac{\partial^2
f(\cdot)}{\partial l\partial m}$ is continuous, as long as
$\ba^H\mRi\ba$ does not vanish. Indeed $\ba\neq \bf{0}$ and $\mRi$
is positive definite (since $\mR$ is positive definite), so
$\ba^H\mRi\ba\neq0$. Therefore, $\frac{\partial^2 f(\cdot)}{\partial
l\partial m}$ is continuous. Similarly, $\frac{\partial^2
f(\cdot)}{\partial m\partial l}$ is continuous. Hence, $
\frac{\partial^2 f(\cdot)}{\partial l\partial m} = \frac{\partial^2
f(\cdot)}{\partial m\partial l}$. This completes our proof.
$\indent\square$
\end{document}